\documentclass{aastex63}

\usepackage{amsmath}
\usepackage{nicefrac}
\usepackage{bbm}

\DeclareMathOperator*{\argmin}{arg\,min}
\DeclareMathOperator*{\argmax}{arg\,max}
\DeclareMathOperator{\asinh}{sinh^{-1}}

\usepackage{CJK} 

\received{*}
\revised{*}
\accepted{*}
\submitjournal{*}

\shorttitle{Deep Potential - Method}
\shortauthors{Green et al.}
\graphicspath{{./}{figures/}}

\begin{document}
\begin{CJK*}{UTF8}{gbsn}

\title{Deep Potential: Recovering the gravitational potential from a snapshot of phase space}

\correspondingauthor{Gregory M. Green}
\email{gregorymgreen@gmail.com}

\author[0000-0001-5417-2260]{Gregory M. Green}
\affiliation{Max Planck Institute for Astronomy\\
  K\"{o}nigstuhl 17, D-69117 Heidelberg, Germany}

\author[0000-0001-5082-9536]{Yuan-Sen Ting (丁源森)}
\affiliation{Research School of Astronomy \& Astrophysics, Australian National University\\
  Cotter Rd., Weston, ACT 2611, Australia}
\affiliation{Research School of Computer Science, Australian National University\\
  Acton, ACT 2601, Australia}

\author[0000-0001-5625-5342]{Harshil Kamdar}
\affiliation{Harvard University, Department of Astronomy\\
  60 Garden St., Cambridge, MA 02138, United States}



\begin{abstract}
  One of the major goals of the field of Milky Way dynamics is to recover the gravitational potential field. Mapping the potential would allow us to determine the spatial distribution of matter -- both baryonic and dark -- throughout the Galaxy. We present a novel method for determining the gravitational field from a snapshot of the phase-space positions of stars, based only on minimal physical assumptions, which makes use of recently developed tools from the field of deep learning. We first train a normalizing flow on a sample of observed six-dimensional phase-space coordinates of stars, obtaining a smooth, differentiable approximation of the distribution function. Using the Collisionless Boltzmann Equation, we then find the gravitational potential -- represented by a feed-forward neural network -- that renders this distribution function stationary. This method, which we term ``Deep Potential,'' is more flexible than previous parametric methods, which fit restricted classes of analytic models of the distribution function and potential to the data. We demonstrate Deep Potential on mock datasets, and demonstrate its robustness under various non-ideal conditions. Deep Potential is a promising approach to mapping the density of the Milky Way and other stellar systems, using rich datasets of stellar positions and kinematics now being provided by \textit{Gaia} and ground-based spectroscopic surveys.
\end{abstract}

\keywords{Milky Way dynamics (1051), Stellar dynamics (1596), Neural networks (1933), Gravitational fields (667), Astrostatistics (1882)}


\section{Introduction}

To know the gravitational potential of the Milky Way is to know the three-dimensional distribution of matter. Stars and gas make up most of the baryonic mass of the Galaxy. However, thus far, dark matter is only detectable through its gravitational influence. Mapping the gravitational potential in 3D is therefore key to mapping the distribution of matter -- both baryonic and dark -- throughout the Galaxy.

The trajectories of stars orbiting in the Milky Way are guided by gravitational forces. If it were possible to directly measure accelerations of individual stars due to the Galaxy's gravitational field, then each star's acceleration would indicate the local gradient of the gravitational potential \citep{Quercellini2008,Ravi2019,Chakrabarti2020GalacticAcceleration}. However, the typical scale of these gravitational accelerations -- on the order of $1\,\mathrm{cm\,s^{-1}\,yr^{-1}}$ -- is well below the precision achieved by current spectroscopic and astrometric instruments \citep{Silverwood2019}.

In general, we are only able to observe a frozen snapshot of stellar positions and velocities. As the gravitational potential guides the motion of stars through phase space, it determines how the phase-space density, the ``distribution function,'' evolves in time. Unless one invokes further assumptions, any gravitational potential is consistent with any snapshot of the distribution function, as the potential only determines the \textit{time evolution} of the distribution function. A critical assumption of most dynamical modeling of the Milky Way is therefore that the Galaxy is in a steady state, meaning that its distribution function does not vary in time \citep{Binney2013,BlandHawthornGerhard2016}.

State-of-the-art dynamics modeling techniques generally work with simplified analytic models of the distribution function and gravitational potential. The results produced by such techniques can only be as good as the models that are assumed. Astrometric and spectroscopic surveys, such as \textit{Gaia} \citep{Prusti2016}, LAMOST \citep{Cui2012LAMOST,Zhao2012LAMOST}, GALAH \citep{DeSilva2015GALAH}, APOGEE \citep{Majewski2017APOGEE} and SDSS-V \citep{Kollmeier2017}, are vastly expanding the number of stars with measured phase-space coordinates, revealing that the distribution function of the Milky Way is richly structured \citep{Antoja2018Nature,Trick2019DR2Actions}. This motivates us to go beyond simple parametric models. Here, we demonstrate a technique that learns highly flexible representations of both the distribution function and potential, capable of capturing the complexity of newly available data. Our method makes only minimal assumptions about the underlying physics:
\begin{enumerate}
    \item Stars orbit in a gravitational potential, $\Phi \left( \vec{x} \right)$.
    \item We have observed the phase-space coordinates of a population of stars that are statistically stationary (\textit{i.e.}, whose phase-space distribution does not change in time).
    \item Matter density is non-negative everywhere. The gravitational potential is related to the matter density, ${\rho \left( \vec{x} \right)}$, by Poisson's equation: ${\nabla^2 \Phi = 4\pi G \rho}$. Thus, ${\nabla^2 \Phi \geq 0}$.
\end{enumerate}
We do not require that the mass density $\rho \left( \vec{x} \right)$ that generates the gravitational potential be consistent with the stellar population that we use as kinematic tracers. That is, the density $\rho \left( \vec{x} \right)$ may contain additional components (such as dark matter) that go beyond our kinematic tracer population.

We make use of a number of mathematical tools from the field of deep learning. We represent the distribution function using a normalizing flow, a highly flexible representation of a probability density function \citep{Rezende2015NormalizingFlows,Kobyzev2019NormalizingFlows,Papamakarios2019NormalizingFlows}. We represent the gravitational potential using a densely connected feed-forward neural network (sometimes called a ``multi-layer perceptron,'' or MLP; see \citealt{Rosenblatt1961MLP,Rumelhart1986Backprop,LeCun2012Backprop}), which takes a 3D position vector and outputs a scalar. These mathematical tools are particularly useful for dynamical modeling, as they can be automatically differentiated using backpropagation (for an introduction to automatic differentiation, see \citealt{baydin-2018-ad-machinelearning}). We can thus easily calculate derivatives that commonly appear in the fundamental equations of gravitational dynamics, such as the Laplacian of the potential or the phase-space gradients of the distribution function.

Our method for recovering the potential, which we term ``Deep Potential,'' can be summarized as follows: we first train a normalizing flow to represent the distribution of the observed phase-space coordinates of the stars, and then find the gravitational potential that renders this distribution stationary, subject to the constraint that matter density must be positive.

This paper builds on the short workshop paper \citet{GreenTing2020NeurIPS}, providing a more pedagogical explanation of Deep Potential (Section~\ref{sec:method}), demonstrating its performance on toy systems with spherical symmetry and axisymmetry (Section~\ref{sec:demonstrations}), and testing its performance under various non-ideal conditions: observational errors, selection functions and non-stationarity (Section~\ref{sec:non-ideal}). Finally, we discuss prospects for further development of Deep Potential and its application to real data (Section~\ref{sec:discussion}).

\section{Method}
\label{sec:method}

\begin{figure*}
  \centering
  \includegraphics[width=0.8\linewidth]{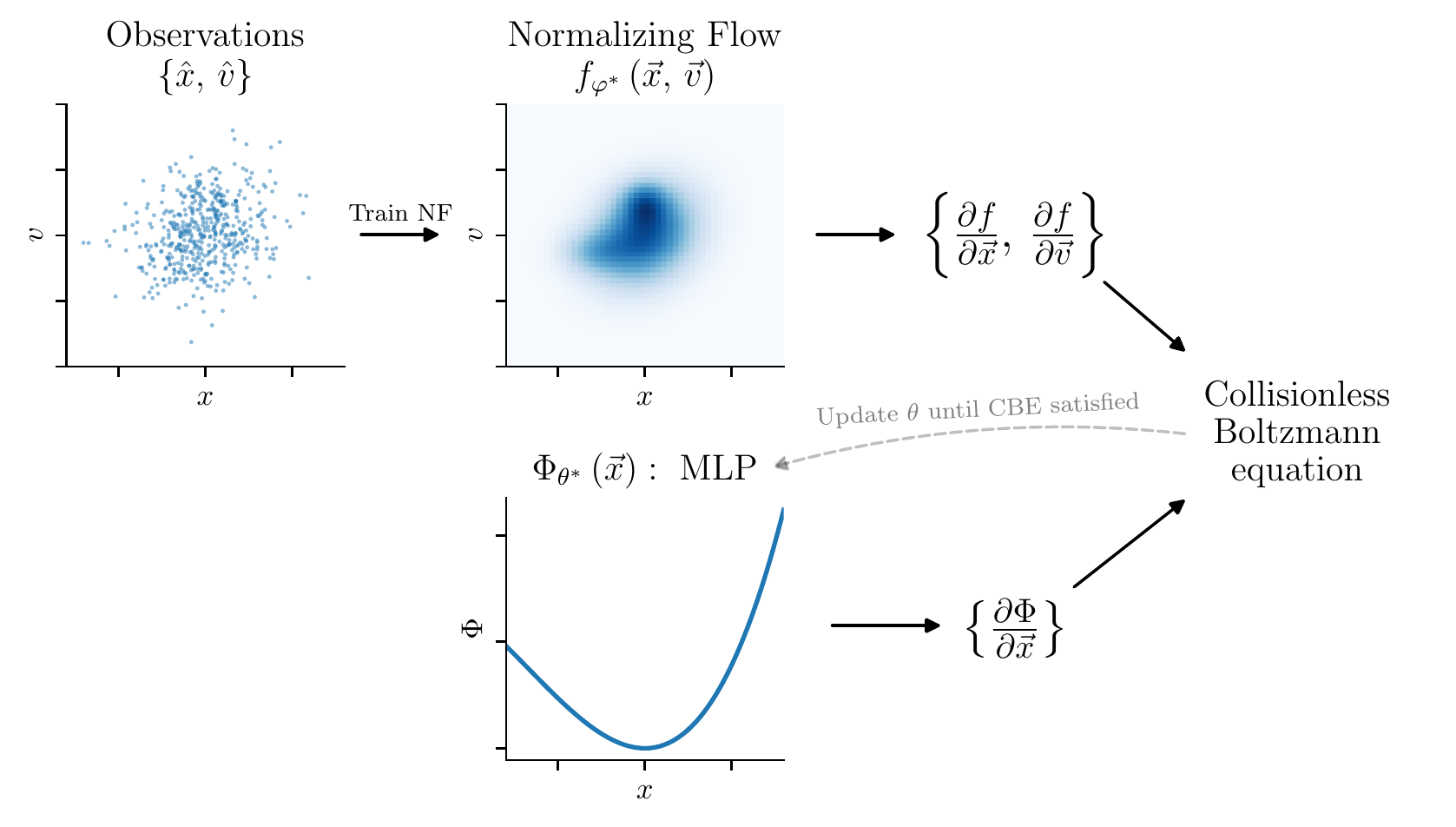}
  \caption{Overview of our method. Using observed phase-space information, we train a normalizing flow to represent the distribution function, $f \left( \vec{x} ,\, \vec{v} \right)$. We then calculate the gradients of $f$ at a large number of points in phase space. We represent the gravitational potential, $\Phi \left( \vec{x} \right)$, by a feed-forward neural network. We update the neural network until the gradients of the potential and the distribution function satisfy the Collisionless Boltzmann Equation (CBE) for a stationary system at the sampled points in phase space. We heavily penalize solutions for which $\nabla^2 \Phi < 0$, which would correspond to negative matter densities.}
  \label{fig:overview}
\end{figure*}

Our first assumption is that stars orbit in a background gravitational potential, $\Phi \left( \vec{x} \right)$. The density of an ensemble of stars in six-dimensional phase space (position $\vec{x}$ and velocity $\vec{v}$) is referred to as the \textit{distribution function}, $f \! \left( \vec{x} , \vec{v} \right)$. Liouville's theorem states that the total derivative of the distribution function of a collisionless system (in which the particles are not scattered by close-range interactions) is zero along the orbits of the particles. For an ensemble of particles orbiting in a gravitational potential, this leads to the Collisionless Boltzmann Equation:
\begin{align}
  \frac{\mathrm{d}f}{\mathrm{d}t}
  =
  \frac{\partial f}{\partial t}
  \ + \!\!\!\!
  \sum_{\mathrm{dimension}\ i} \!
  \left(
    v_i \, \frac{\partial f}{\partial x_i}
    -\frac{\partial \Phi}{\partial x_i} \frac{\partial f}{\partial v_i}
  \right)
  = 0 \, .
  \label{eqn:f-total-derivative}
\end{align}
Our second assumption, that the distribution function is stationary, implies that the density in any given region of phase space is constant in time: $\frac{\partial f}{\partial t} = 0$. This assumption links gradients of the distribution function to gradients of the gravitational potential:
\begin{align}
  \frac{\partial f}{\partial t}
  =
  \sum_{i} \!
  \left(
    \frac{\partial \Phi}{\partial x_i} \frac{\partial f}{\partial v_i}
    -
    v_i \, \frac{\partial f}{\partial x_i}
  \right)
  = 0 \, .
  \label{eqn:stationarity}
\end{align}
\textit{Once we can describe the distribution function of a stationary system, in all physically plausible cases, the gravitational potential can be uniquely determined (up to an additive constant) by solving the Collisionless Boltzmann Equation} (see Appendix~\ref{app:existence-uniqueness}, as well as \citealt{An2021Uniqueness}). Realistic physical systems will not be completely stationary, and as such, there may not exist any potential which would render the system stationary. In general, therefore, Deep Potential recovers the potential which \textit{minimizes} the amount of non-stationarity in the system (using a measure that will be discussed below). Fig.~\ref{fig:overview} gives a graphical overview of Deep Potential.

Note that we do not assume that the gravitational potential is sourced by the observed stellar population alone. Accordingly, we do not impose the condition
\begin{align}
  \nabla^2 \Phi = 4 \pi G \int \!f \left( \vec{x} , \vec{v} \right) \mathrm{d}^3\vec{v} \, .
\end{align}
Additional mass components beyond the observed stellar population, such as unobserved stars or dark matter, may also contribute to the gravitational field. It is even possible, within our formalism, for the mass density corresponding to the recovered gravitational potential to be less than the mass density of the observed tracer population. We treat the observed stars merely as test masses, whose kinematics we can use to map the background potential.

\subsection{Modeling the distribution function}

In practice, when we observe stellar populations, we obtain a discrete sample of points in phase space, which we will refer to as $\left\{ \hat{x} , \hat{v} \right\}$. We do not directly observe the smooth distribution function from which the points are drawn, $f \left( \vec{x} , \vec{v} \right)$. Typical methods of surmounting this difficulty include fitting a simple parametric model of the distribution function to the observed sample of phase-space locations of stars, often in action space \citep[e.g.,][]{Binney2010DF,TingRixBovyVandeVen2013,BovyRix2013,McMillanBinney2013,Piffl2014,Trick2016}, or to work with moments of the distribution function \citep[``Jeans modeling,'' e.g.,][]{Gnedin2010MassProfile,HagenHelmi2018VerticalForce}. Here, we instead represent the distribution function using a highly flexible tool from the field of deep learning, known as a ``normalizing flow'' \citep[for a review, see][]{Kobyzev2019NormalizingFlows,Papamakarios2019NormalizingFlows}. A normalizing flow represents an arbitrary probability density function $p \left( \vec{z} \right)$ as a coordinate transformation of a simpler distribution. For example, we might begin with a unit normal distribution in the space $\vec{z}^{\,\prime}$. Given a bijection (i.e., an invertible coordinate transformation) between $\vec{z}$ and $\vec{z}^{\,\prime}$, we can transform this simple distribution in $\vec{z}^{\,\prime}$ into a possibly complicated distribution in $\vec{z}$:
\begin{align}
    p \left( \vec{z} \right)
    = \mathcal{N} \left( \vec{z}^{\,\prime} \mid 0 , \mathbbm{1} \right)
      \left| \frac{\partial \vec{z}^{\,\prime}}{\partial \vec{z}} \right|
    \, .
\end{align}
The key is to find a parameterized class of highly flexible, nonlinear, bijective transformations, for which the Jacobian can be efficiently calculated. In this paper, we call the parameters governing this family of coordinate transformations $\varphi$, and we refer to the resulting probability density function in $\vec{z}$ as $p_{\varphi} \left( \vec{z} \right)$. Given a set of points $\left\{ \vec{z} \right\}$ that are drawn from an unknown distribution $p \left( \vec{z} \right)$, we can then search for the parameters $\varphi$ that maximize the likelihood of the points. This then yields a smooth approximation of the distribution from which the points $\left\{ \vec{z} \right\}$ were drawn. The ``training'' of the normalizing flow usually proceeds through stochastic gradient descent using batches of points. There are several classes of bijective transformations commonly used in normalizing flows -- an overview can be found in \citet{Kobyzev2019NormalizingFlows}. In this work, we use FFJORD \citep{Grathwohl2018FFJORD}, though the detailed choice of which class of transformations to use is not critical to our method. Research into normalizing flows is proceeding rapidly, and as better methods become available, they can be incorporated into our method as a drop-in replacement. We give a more detailed description of our normalizing flow implementation in Section~\ref{sec:implementation}.

In order to obtain a smooth approximation to the distribution function, we train a normalizing flow on an ensemble of stars with measured phase-space information. We refer to the normalizing flow as $f_{\varphi} \! \left( \vec{x} , \vec{v} \, \right)$, where $\varphi$ refers to the parameters controlling the coordinate transformation in the flow. The best-fit flow parameters are those that maximize the likelihood of the phase-space coordinates of the stars:
\begin{align}
  \varphi^{\ast}
  &=
  \argmax_{\varphi} \bigg[
    \ln p \left( \left\{ \hat{x} , \hat{v} \right\} \mid \varphi \right)
  \bigg]
  \\
  &=
  \argmax_{\varphi}
  \Bigg[
    \sum_{\mathrm{star}\ k}
      \ln f_{\varphi} \! \left( \hat{x}_k , \hat{v}_k \right)
  \Bigg]
  \, .
  \label{eqn:df-best-fit}
\end{align}
We thus obtain an approximation to the distribution function, $f_{\varphi^{\ast}}$. The great advantage of using a normalizing flow is that our representation is both highly flexible and auto-differentiable. When implemented in a standard deep learning framework, such as Tensorflow \citep{tensorflow2015-whitepaper} or PyTorch \citep{Paszke2019PyTorch}, it is possible to automatically differentiate the distribution function at arbitrary points in phase space, in order to obtain the terms $\nicefrac{\partial f}{\partial \vec{x}}$ and $\nicefrac{\partial f}{\partial \vec{v}}$ in the Collisionless Boltzmann Equation.

\subsection{Modeling the gravitational potential}

After learning the distribution function, we find the gravitational potential $\Phi \left( \vec{x} \right)$ that renders the distribution function stationary. The distribution is stationary when Eq.~\eqref{eqn:stationarity} is satisfied everywhere in phase space. We parameterize the gravitational potential as a feed-forward neural network, which takes a 3-vector, $\vec{x}$, and returns a scalar, $\Phi$. We denote the trainable parameters of this network (i.e., the weights and biases) by $\theta$, and the resulting approximation function as $\Phi_{\theta} \! \left( \vec{x} \right)$. For any given value of $\theta$, we can calculate the non-stationarity of our approximation of the distribution function at any arbitrary point $\left( \vec{x} , \vec{v} \right)$ in phase space:
\begin{align}
  \frac{\partial f_{\varphi^{\ast}}}{\partial t}
  =
  \sum_{i} \!
  \left(
    \frac{\partial \Phi_{\theta}}{\partial x_i}
    \frac{\partial f_{\varphi^{\ast}}}{\partial v_i}
    -
    v_i \, \frac{\partial f_{\varphi^{\ast}}}{\partial x_i}
  \right)
  \, .
\end{align}
This is essentially a variational method, in which our Ansatz for the gravitational potential is a neural network $\Phi_{\theta}$, and in which we vary the parameters $\theta$ to minimize non-stationarity of the distribution function.

We require a measure of the non-stationarity of the distribution function throughout all of phase space, which we can then use to find the optimal gravitational potential. We also require that the matter density be non-negative everywhere in space. By Poisson's equation, which links the potential to the density, this implies that $\nabla^2 \Phi \geq 0$. Rather than strictly enforcing this condition, we will penalize solutions that include negative matter densities. There are different ways that one could quantify the total non-stationarity and extent of negative matter density of a system, but we find it useful to consider measures that have the following form:
\begin{align}
    L \left( \theta \right) =
    \int
    \mathcal{L} \left(
        \frac{\partial f \left( \vec{x}, \vec{v} \right)}{\partial t} ,\,
        \nabla^2 \Phi \left( \vec{x} \right)
    \right)
    W \left( \vec{x}, \vec{v} \right)
    \mathrm{d}^3 \vec{x} \, \mathrm{d}^3 \vec{v}
    \, ,
\end{align}
where $\mathcal{L}$ is a function that takes the raw values of $\nicefrac{\partial f}{\partial t}$ and $\nabla^2 \Phi$ (both of which depend on the potential, and thus on the parameters $\theta$) at a given point and returns a positive scalar penalty, and $W$ determines how different points in phase space are weighted. In this work, we choose
\begin{align}
    \mathcal{L} 
    &=
    \asinh \! \left(
      \alpha
      \left| \frac{\partial f}{\partial t} \right|
    \right)
    \! + \!
    \lambda \, \asinh \! \left(
      \beta
      \max \left\{
        -\nabla^2 \Phi_{\theta}
        ,
        0
      \right\}
    \right)
    \, .
\end{align}
The first term penalizes non-stationarity, while the second term penalizes negative masses. We first take the absolute value of $\nicefrac{\partial f}{\partial t}$, in order to penalize positive and negative changes in the phase-space density equally. The inverse hyperbolic sine function down-weights large values, while the constant $\alpha$ sets the level of non-stationarity at which our penalty transitions from being approximately linear to being approximately logarithmic. This down-weighting of large non-stationarities is not strictly necessary, but in our experiments leads to more consistent results. In our toy models, typical distribution-function gradients can vary by orders of magnitude between different regions of phase space. Down-weighting large non-stationarities prevents small regions of phase-space with large distribution-function gradients from dominating the fit. The second term penalizes negative mass densities, again using the inverse hyperbolic sine to down-weight large values. The constant $\beta$ sets the negative mass density at which this penalty transitions from linear to logarithmic. The hyperparameter $\lambda$ sets the relative weight given to the non-stationarity and negative-mass penalties (we choose $\lambda = 1$ in this work). We set the weight function $W \left( \vec{x}, \vec{y} \right)$ to be equal to our approximation of the distribution function. Thus,
\begin{align}
    L =
    \int
    \mathcal{L} \left( \vec{x}, \vec{v}, \theta \right)
    f_{\varphi^{\ast}} \! \left( \vec{x}, \vec{v} \right)
    \mathrm{d}^3 \vec{x} \, \mathrm{d}^3 \vec{v}
    \, .
\end{align}
This integral is computationally expensive to evaluate directly, but it can be approximated by averaging the value of the bracketed term over samples in phase space drawn from the distribution function:
\begin{align}
  L \left( \theta \right) \approx
  \left<
    \mathcal{L} \left( \vec{x}, \vec{v}, \theta \right)
  \right>_{
    \vec{x} , \vec{v} \, \sim \, f_{\varphi^{\ast}}
  }
  \, .
  \label{eqn:theta-loss}
\end{align}
We find the gravitational potential parameters, $\theta^{\ast}$, that minimize the above loss function:
\begin{align}
  \theta^{\ast}
  &=
  \argmin_{\theta} L \left( \theta \right)
  \, .
  \label{eqn:theta-best-fit}
\end{align}
Finally, in order to regularize the potential and prevent over-fitting, it is possible to add a term to $L \left( \theta \right)$ that penalizes complicated potentials. As we implement $\Phi_{\theta} \left(\vec{x}\right)$ as a neural network, one obvious possibility is to penalize large weights in the network. For example, one might choose to add an $L_2$ weight regularization penalty to the logarithm of the loss:
\begin{align}
    \ln L \left(\theta\right) + \frac{\ell_2}{N_w} \sum_{i=1}^{N_w} w_i^2 \, ,
    \label{eqn:penalized-loss}
\end{align}
where $\left\{w\right\}$ are the neural-network weights contained in the parameters $\theta$, $N_w$ is the number of weights, and $\ell_2$ is a small constant that sets the strength of the penalty.

In order to optimize our model of the potential, we begin by drawing a large number of phase-space points from our approximation to the distribution function, and calculating the gradients $\nicefrac{\partial f_{\varphi^{\ast}}}{\partial \vec{x}}$ and $\nicefrac{\partial f_{\varphi^{\ast}}}{\partial \vec{v}}$ at each point. Given a value of $\theta$, this allows us to calculate $\nicefrac{\partial f_{\varphi^{\ast}}}{\partial t}$ at any of these points in phase space. We randomly initialize the gravitational potential parameters, $\theta$, and then use stochastic gradient descent on batches of our phase-space points to minimize the loss and obtain $\theta^{\ast}$.

Once the potential is obtained, forces and densities can be computed from its gradient and Laplacian, respectively. Because the potential is represented by a neural network, these derivatives can be calculated using auto-differentiation.

\subsection{Additional considerations}
\label{sec:additional-considerations}

Before discussing our implementation of Deep Potential, we offer a few considerations on the approach sketched out above.

First, our stationarity assumption is \textit{frame-dependent}. In the formulation we have given above, the frame of reference must be fixed at the outset (for example, one might choose a frame that is at rest with respect to the Galactic Center, or which is rotating with the pattern speed of the Galactic bar). However, it is possible to generalize our method by assuming that the distribution function is stationary in an arbitrary reference frame that is moving (or rotating) with respect to the frame in which the positions and velocities of the kinematic tracers have been measured. Instead of asserting that $\partial f / \partial t = 0$, we can assert that the distribution function appears stationary for observers following a certain set of trajectories, defined by a spatially dependent velocity field $\vec{u} \left( \vec{x} \right)$. This leads to the condition,
\begin{align}
  \frac{\partial f}{\partial t} + \sum_i \left( u_i \frac{\partial f}{\partial x_i} + \dot{u}_i \frac{\partial f}{\partial v_i} \right) = 0 \, ,
\end{align}
where $\dot{\vec{u}}$ is the acceleration along the trajectory. Inserting this condition into the Collisionless Boltzmann Equation (Eq.~\ref{eqn:f-total-derivative}), we obtain a modified version of our stationarity condition (Eq.~\ref{eqn:stationarity}):
\begin{align}
  \sum_{i} \!
  \left[
    \left( \frac{\partial \Phi}{\partial x_i} + \dot{u}_i \right) \frac{\partial f}{\partial v_i}
    -
    \left( v_i - u_i \right) \, \frac{\partial f}{\partial x_i}
  \right]
  = 0 \, .
  \label{eqn:stationarity-transformed-frame}
\end{align}
The stationarity condition is nearly identical to Eq~\eqref{eqn:stationarity}, except for two modifications: a velocity shift ($\vec{v} \rightarrow \vec{v} - \vec{u}$), and a shift in the potential gradients ($\nabla \Phi \rightarrow \nabla \Phi + \dot{\vec{u}}$) related to fictitious forces felt by an accelerating observer.

The form of the velocity field $\vec{u} \left( \vec{x} \right)$ determines the change of reference frame. A frame of reference that is moving with constant speed with respect to the standard coordinates $\left( \vec{x}, \vec{v} \right)$ is defined by a constant $\vec{u} \left( \vec{x} \right) = \vec{u}_0$, with time derivative $\dot{\vec{u}} = 0$. A frame of reference that is rotating with angular velocity vector $\vec{\Omega}$ about an axis passing through the point $\vec{x}_0$ is described by $\vec{u} = \vec{\Omega} \times \left( \vec{x} - \vec{x}_0 \right)$, with time derivative $\dot{\vec{u}} = \vec{\Omega} \times \vec{u} = \vec{\Omega} \times \left[ \vec{\Omega} \times \left( \vec{x} - \vec{x}_0 \right) \right]$. The parameters that define the change of frame, such as $\vec{u}_0$, $\vec{x}_0$ and $\vec{\Omega}$, can be either fixed or treated as free parameters. If they are treated as free parameters, they can be varied at the same time as the potential is fit, so as to find the reference frame in which non-stationarity is minimized. The Ansatz is then that the system is stationary \textit{in some reference frame that is not determined at the outset}, but which is related to the measurement frame by a simple transformation.

A second consideration is that we assume that the kinematic tracers are observed in all six phase-space dimensions. While it is possible to obtain full phase-space information for millions of stars in the Milky Way, for many extragalactic systems, one or more phase-space dimensions (such as line-of-sight position within the system or proper motion) are missing. One approach to dealing with missing phase-space dimensions would be to fit the distribution function in the lower-dimensional space, and then to recover the missing dimensions with symmetry assumptions (e.g., in the absence of proper motions, assume that velocities are isotropically distributed).

Third, we do not impose any symmetries on the gravitational potential. However, it is possible to impose simple symmetries. For example, one could impose axisymmetry on the potential by only feeding cylindrical radius, $R$, and height, $z$, into the neural network that represents it: $\Phi \left( R,z \right)$. It is also possible to \textit{favor} -- without strictly imposing -- symmetries. For example, one could favor axisymmetry by representing the potential as a sum of an axisymmetric component and a regularized component without axisymmetry.

We leave deeper consideration of these issues to future work, and focus here on demonstrating Deep Potential with full phase-space data in a fixed frame of reference and no assumed symmetries in the potential.

\subsection{Implementation}
\label{sec:implementation}

In this paper, we implement Deep Potential in Tensorflow 2. All of our code is publicly available, under a permissive license that allows re-use and modification with attribution.\footnote{Our code is available at \url{https://github.com/gregreen/deep-potential}. A PyTorch implementation of Deep Potential, using a different normalizing flow architecture (a Neural Spine Flow; \citealt{Durkan2019NeuralSplineFlows}), is available at \url{https://github.com/tingyuansen/deep-potential}.}

To represent the distribution function, we use a chain of three FFJORD normalizing flows \citep{Grathwohl2018FFJORD}, each with 3 densely connected hidden layers of 128 neurons and a $\tanh$ activation function. For our base distribution, we use a multivariate Gaussian distribution with mean and variance along each dimension set to match the training dataset.\footnote{Another, nearly equivalent option is to "standardize" the coordinates before feeding them into the first FFJORD bijection, by shifting and scaling so that the entire training dataset has zero mean and unit variance along each dimension.} During training, we impose Jacobian and kinetic regularization with strength $10^{-4}$ \citep{Finlay2020FFJORDRegularization}, which penalizes overly complex flow models and tends to reduce training time. We train our flows using the Rectified Adam optimizer \citep{Liu2019RAdam}, with a batch size of $2^{13}$ (8096). We find that this relatively large batch size leads to faster convergence (in wall time) than more typical, smaller batch sizes. We begin the training with a ``warm-up'' phase that lasts 2048 steps, in which the learning rate linearly increases from 0 to 0.005. Thereafter, we use a constant learning rate. We decrease the learning rate by a factor of two whenever the training loss fails to decrease 0.01 below its previous minimum for 2048 consecutive steps (this period is termed the ``patience''). We terminate the training when the loss ceases to decrease by 0.01 over the course of several consecutive drops in learning rate. In our experiments with mock datasets (see Section~\ref{sec:demonstrations}), these settings allow us to accurately recover a range of different distribution functions with varying levels of complexity (e.g., with symmetries or caustics).

After training our normalizing flow, we draw $2^{20}$ ($\sim$1~million) phase-space coordinates, and calculate the gradients $\partial f_{\varphi^{\ast}} / \partial{\vec{x}}$ and $\partial f_{\varphi^{\ast}} / \partial{\vec{v}}$ at each point (using auto-differentiation), for use in learning the gravitational potential.

We represent the gravitational potential using a feed-forward neural network with 4 densely connected hidden layers, each with 512 neurons and a $\tanh$ activation function (we eschew more commonly used activation functions, such as ReLU, which have discontinuous derivatives, as these may lead to unphysical potentials). The network takes a 3-dimensional input (the position $\vec{x}$ in space), and produces a scalar output (the potential). We add in an $L_2$ loss (see Eq.~\ref{eqn:penalized-loss}) with strength $\ell_2 = 0.1$. We train the network using the Rectified Adam optimizer, with batches of $2^{14}$ (16384) phase-space coordinates. We use a similar learning-rate scheme as before, with a warm-up phase lasting 2048 steps, an initial learning rate of 0.001, and a patience of 2048 steps. We set the parameters $\alpha$, $\beta$ and $\lambda$ in Eq.~\eqref{eqn:theta-best-fit}, which affect the penalties on non-stationarity and negative gravitational mass densities, to unity.

When fitting both the distribution function and the gravitational potential, we reserve 25\% of our input data as a validation set. After each training step, we calculate the loss on a batch of validation data, in order to identify possible overfitting to the training data. Such overfitting would manifest itself as a significantly lower training loss than validation loss. In the experiments in this paper, no significant overfitting is observed  -- the difference in the likelihoods of the training and validation sets is typically less than 1\%.

The choices made here are by no means set in stone, and can be altered without changing the overall Deep Potential framework. In particular, rapid advances have been made in research into normalizing flows over the preceding years. As more accurate and/or computationally efficient flows are developed, they can be used by Deep Potential.

\section{Demonstrations}
\label{sec:demonstrations}

We now evaluate the performance of our method on a number of toy physical systems in which the potential is known exactly. We will begin with ideal cases, and then in Section~\ref{sec:non-ideal}, we will show how our method performs in the presence of non-stationarity, observational errors and selection functions.

\subsection{Plummer sphere}
\label{sec:plummer-sphere}

\begin{figure*}
  \centering
  \includegraphics[width=0.95\linewidth]{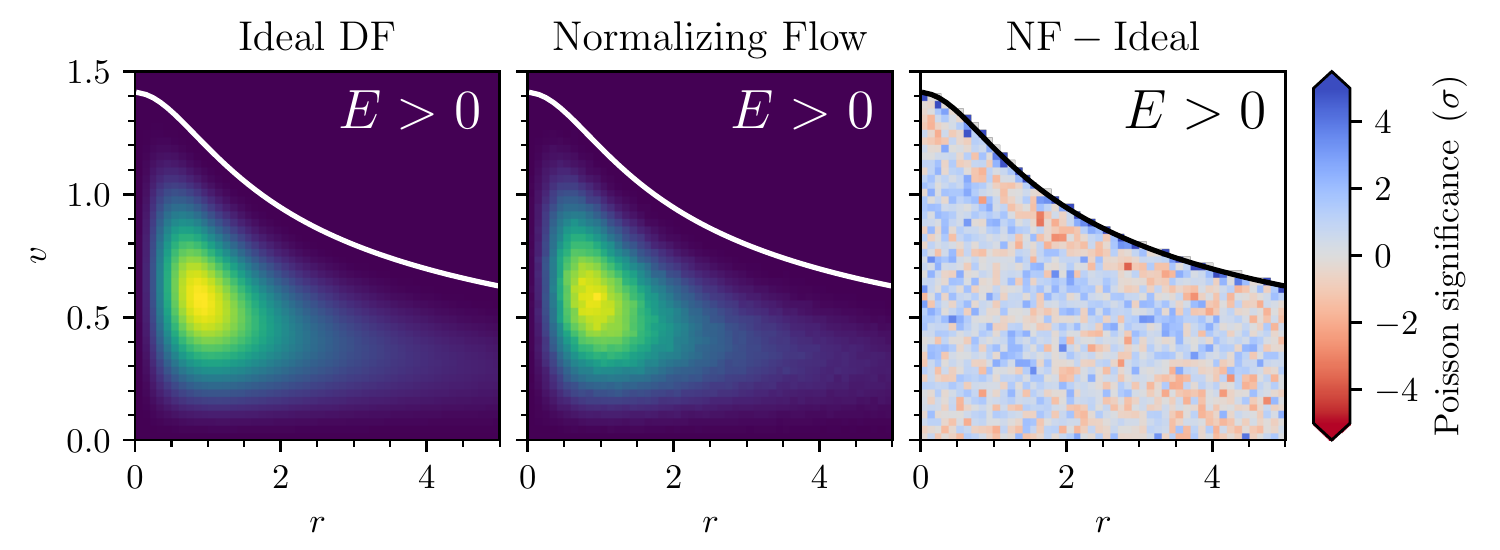}
  \caption{The ideal Plummer sphere distribution function (left panel), a histogram of $2^{20}$ samples drawn from our trained normalizing flow (middle panel), and a comparison of the two (right panel). For the purposes of this figure, we depict phase space in terms of radius and velocity, integrating over the four angular dimensions. The empty region of phase space in which particles would be unbound ($E > 0$) is marked. Our normalizing flow closely matches the true distribution function, as indicated by the small Poisson significances in the right panel. In each bin of $\left( r,v \right)$, we define the Poisson significance as $\left( n_{\mathrm{NF}} - n_{\mathrm{ideal}} \right) / n_{\mathrm{ideal}}^{\nicefrac{1}{2}}$, where $n_{\mathrm{NF}}$ is the number of samples in the bin drawn from the normalizing flow and $n_{\mathrm{ideal}}$ is the number of samples expected from the ideal distribution function.}
  \label{fig:plummer-flow}
\end{figure*}

\begin{figure*}
  \centering
  \includegraphics[width=0.95\linewidth]{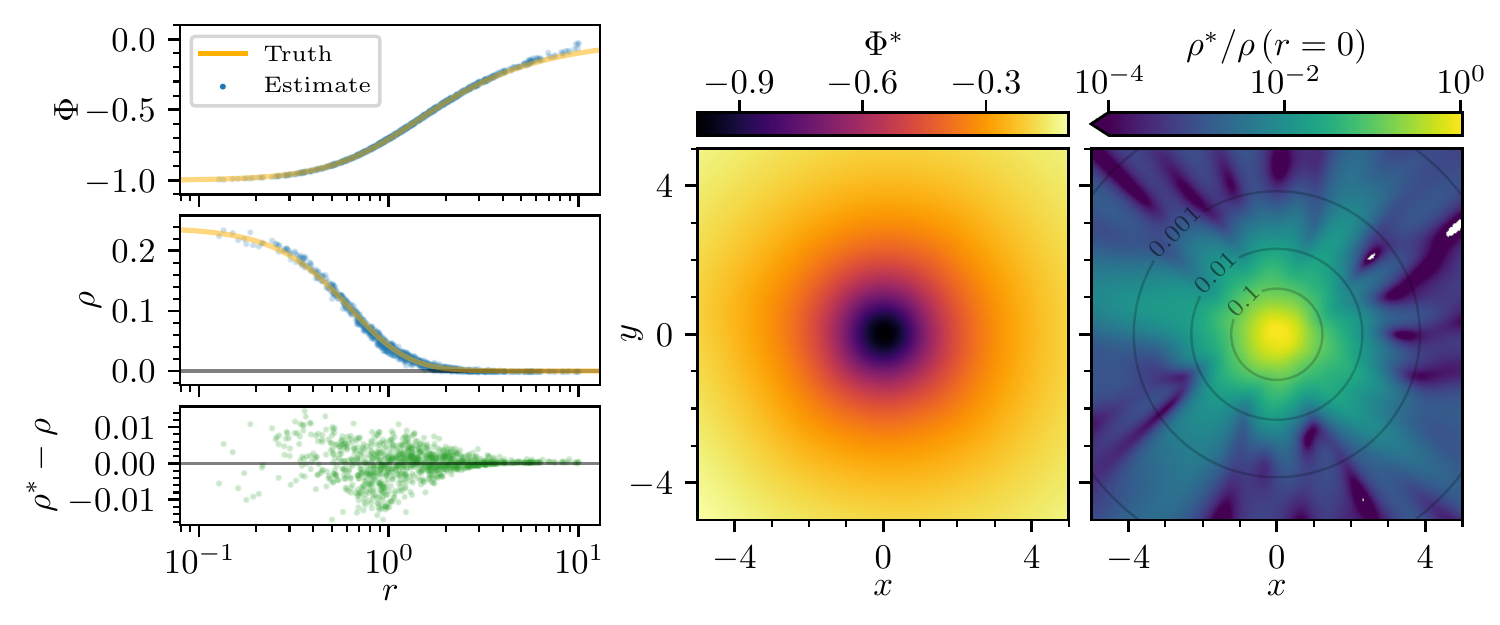}
  \caption{\textbf{Left panels}: Comparison of the theoretical Plummer sphere potential (top) and density (middle) with our result at random positions drawn from the true density profile. The bottom-left panel shows the density residuals (estimates minus truth) at these same points. We accurately recover the true potential and density over a wide range of distances. \textbf{Right two panels}: Our recovered potential (middle) and matter density (right) in the 2D plane defined by $z = 0$. The striped patterns in the recovered density field (in the right panel) are small fluctuations in regions of negligible density, as can be seen from the overplotted contours of true density.}
  \label{fig:plummer-potential}
\end{figure*}

\begin{figure*}
  \centering
  \includegraphics[width=0.85\linewidth]{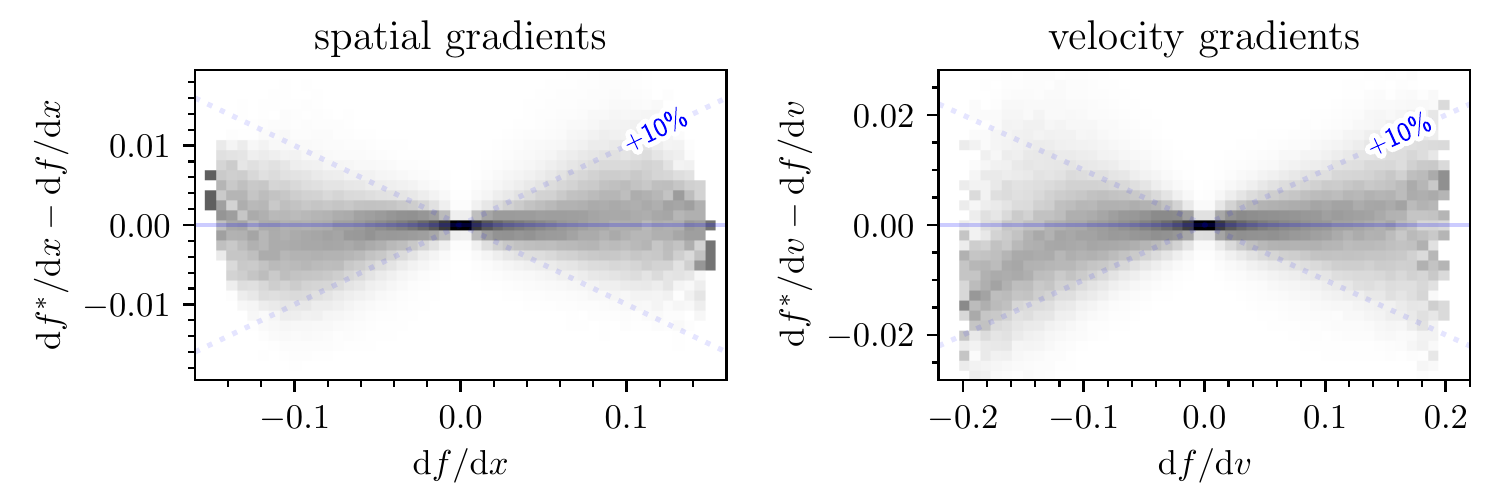}
  \caption{Comparison of the spatial and velocity gradients of our normalizing flow with those of the true Plummer distribution function, at points drawn from the distribution function. Each column is a histogram of the residuals at a given value of $\mathrm{d}f/\mathrm{d}x$ (left panel) or $\mathrm{d}f/\mathrm{d}v$ (right panel). Because the Plummer distribution function is spherically symmetric in space and isotropic in velocity, we group the three spatial gradients together in the left panel, and the three velocity gradients together in the right panel. We generally estimate gradients to better than 10\%, which we find to be sufficient to accurately precisely recover the gravitational potential.}
  \label{fig:plummer-df-gradients}
\end{figure*}

We begin with the Plummer sphere \citep{Plummer1911}, a self-gravitating system with a spherically symmetric density and corresponding gravitational potential, given by
\begin{align}
    \rho \left( r \right) = \frac{3}{4\pi} \left( 1 + r^2 \right)^{-\nicefrac{5}{2}}
    \, ,
    \hspace{0.3cm}
    \Phi \left( r \right) = -\left( 1 + r^2 \right)^{-\nicefrac{1}{2}}
    \, .
\end{align}
The Plummer sphere admits a stationary distribution function with an isotropic velocity distribution, in which the distribution function is only a function of the energy $E$ of the tracer particle (for simplicity, we set mass $m = 1$ for all of the particles):
\begin{align}
    f \! \left( \vec{r}, \vec{v} \right)
    &\propto
    \begin{cases}
      \left[ -E \left( \vec{r}, \vec{v} \right) \right]^{\nicefrac{7}{2}}, & E < 0 \\
      0, & E \geq 0
    \end{cases}
    \, , \notag
    \\ &\hspace{2cm}
    \mathrm{where} \ 
    E = \frac{1}{2} v^2 + \Phi \left( r \right) \, .
\end{align}
We generate mock data by drawing $2^{19}$ (524,288) phase-space coordinates from the above distribution function. We truncate the distribution at a radius of $r = 10$, as in practice, we find that rare particles very far from the origin can sometimes cause large excursions during the training of our normalizing flow.\footnote{Another way to address this issue is to use gradient clipping \citep{Pascanu2012GradientClipping} or similar techniques during training.} Using these coordinates as input data, we first train a normalizing flow to represent the distribution function. Our results are shown in Fig.~\ref{fig:plummer-flow}. We then draw $2^{20}$ (1,048,576) phase-space points from our normalizing flow, and calculate the gradients, $\nicefrac{\partial f_{\varphi^{\ast}}}{\partial \vec{x}}$ and $\nicefrac{\partial f_{\varphi^{\ast}}}{\partial \vec{v}}$, at each point. We use these samples to fit the gravitational potential. Our results are shown in Fig.~\ref{fig:plummer-potential}. We accurately recover both the gravitational potential and the matter density that sources it over a wide range of radii.

We find that the performance of our method depends sensitively on the accuracy with which we recover the gradients of the distribution function. Because the Plummer sphere has an analytic distribution function, we can calculate the true gradients exactly, and compare them with the estimate we obtain from our normalizing flow. As we show in Fig.~\ref{fig:plummer-df-gradients}, we generally recover both spatial and velocity gradients to better than 10\% for our Plummer sphere toy model.

The computational time of our entire method is dominated by the training of the normalizing flow (which can range from a few hours to two days on one Nvidia A100 GPU, depending on the complexity of the distribution function) and the calculation of distribution function gradients at the sampled points. Training the potential takes comparatively little time (less than one hour on a single GPU).

\subsection{Miyamoto-Nagai disk}
\label{sec:miyamoto-nagai}

\begin{figure}
  \centering
  \includegraphics[width=0.48\textwidth]{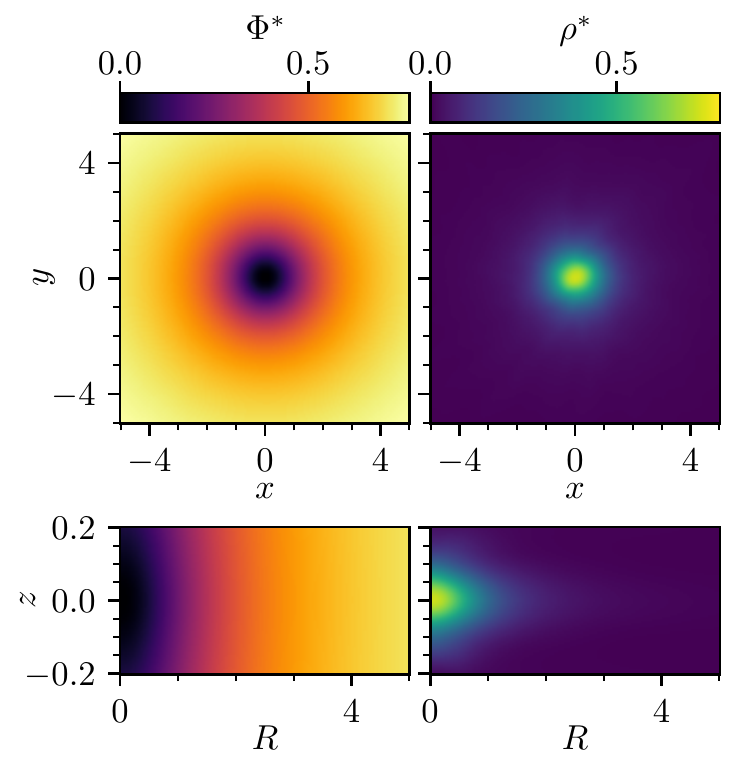}
  \caption{Gravitational potential (left column) and density (right column) recovered in the midplane (upper panels) and the $\left(R,z\right)$-plane (averaged over azimuth, lower panels) of our Miyamoto-Nagai disk system, using ideal mock data.}
  \label{fig:mn-potential-density}
\end{figure}

\begin{figure*}
  \centering
  \includegraphics[width=\linewidth]{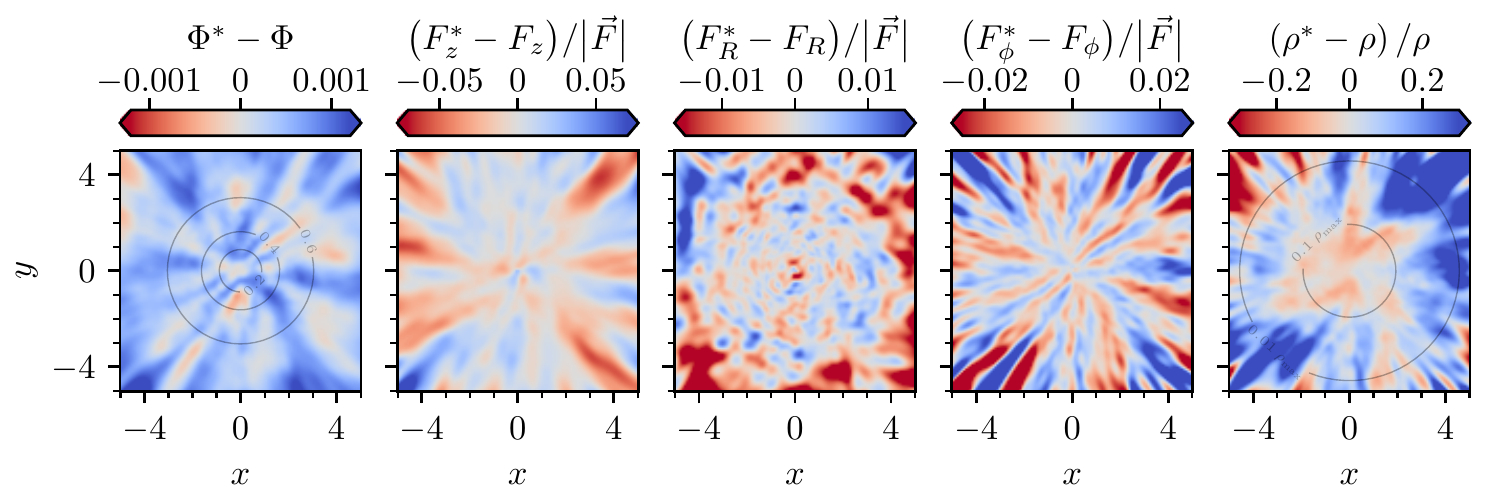}
  \caption{From left to right, residuals (recovered minus ideal) of the potential, $z$-, $R$- and $\phi$-components of force, and density in the midplane ($z = 0$) of our Miyamoto-Nagai disk, as recovered using ideal mock data. The force residuals are divided by the norm of the true force vector at each location, while the density residuals are divided by the true density at each location. In the left and right panels, we overplot potential and density isocontours, respectively.}
  \label{fig:mn-residuals-xy-plane}
\end{figure*}

\begin{figure*}
  \includegraphics[width=0.45\linewidth]{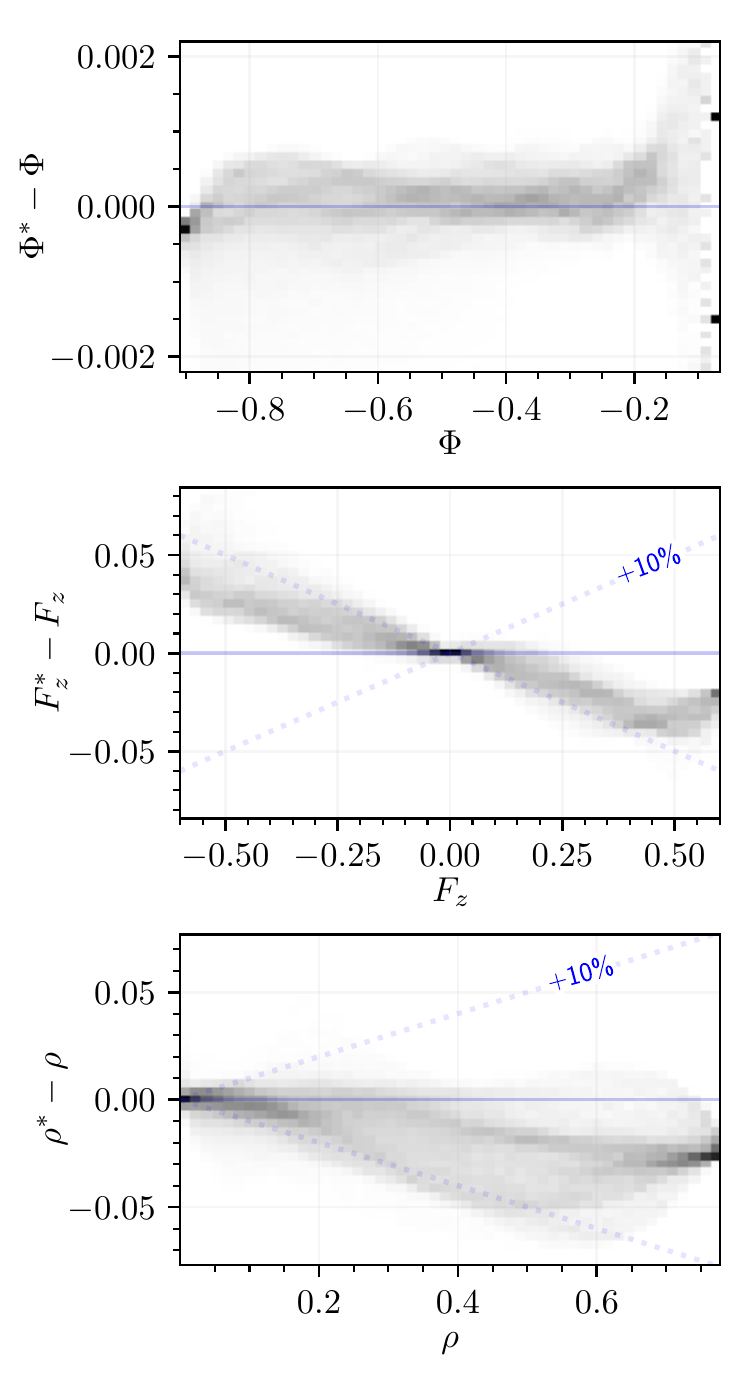}
  \includegraphics[width=0.45\linewidth]{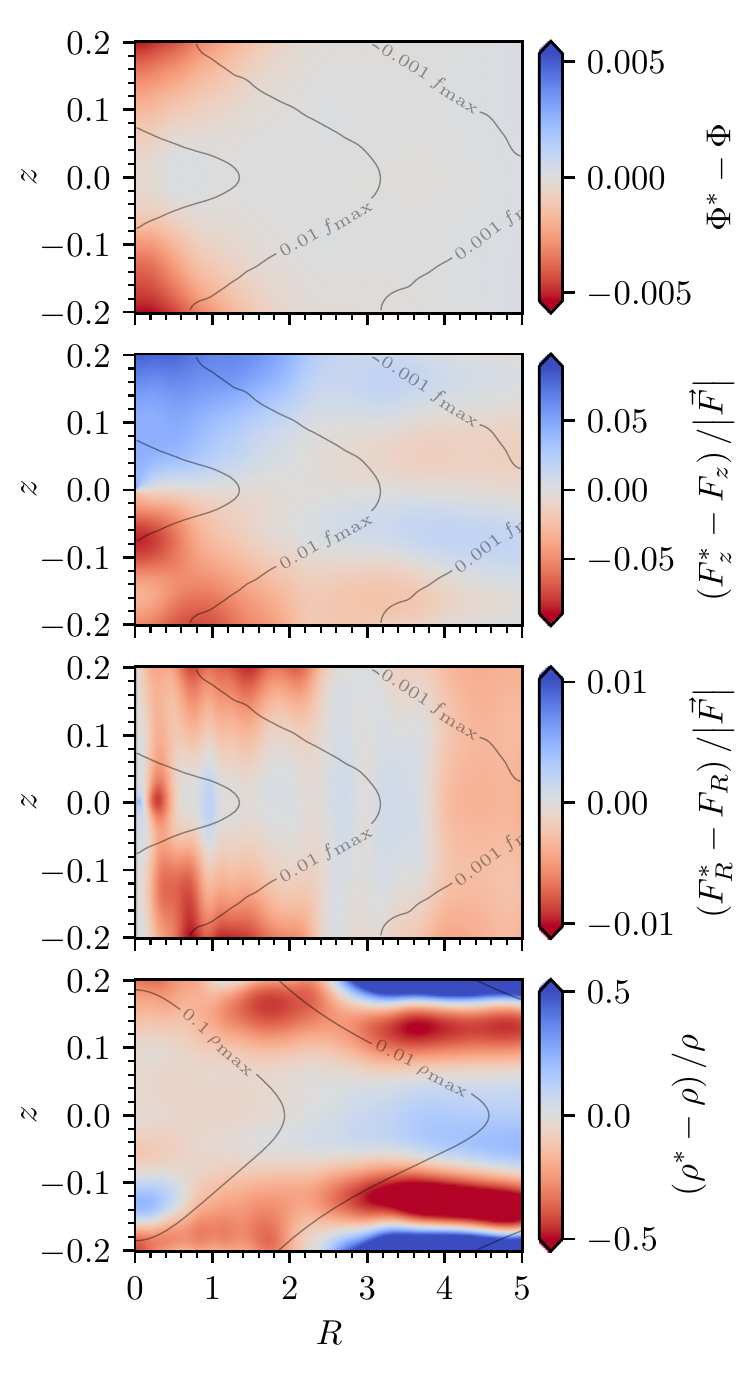}
  \caption{\textbf{Left panels, from top to bottom}: Residuals (recovered minus ideal) of the potential, $z$-component of force, and density of our ideal Miyamoto-Nagai disk, evaluated at the locations of the kinematic tracers. In each bin along the $x$-axis (of potential, force or density), we show a histogram of the residuals. In regions where we have kinematic tracers, we typically recover vertical forces and densities to within 10\%. \textbf{Right panels, from top to bottom}: Residuals of the potential, $z$- and $R$-components of force, and density of our ideal Miyamoto-Nagai disk, averaged over azimuth. In the top three panels on the right, we overplot isocontours of the density of our kinematic tracers, in logarithmic steps of 10$\times$. In the bottom-right panel, we overplot isocontours of the density that sources the gravitational potential ($\tfrac{1}{4\pi G} \nabla^2 \Phi$). Note that in regions of very low density, small \textit{absolute} density residuals lead to large \textit{fractional} density residuals. However, in regions of high density, we generally recover density to within $\sim$10\% of the true value.}
  \label{fig:mn-residuals-tracers-Rz}
\end{figure*}

\begin{figure*}
  \centering
  \includegraphics[width=\linewidth]{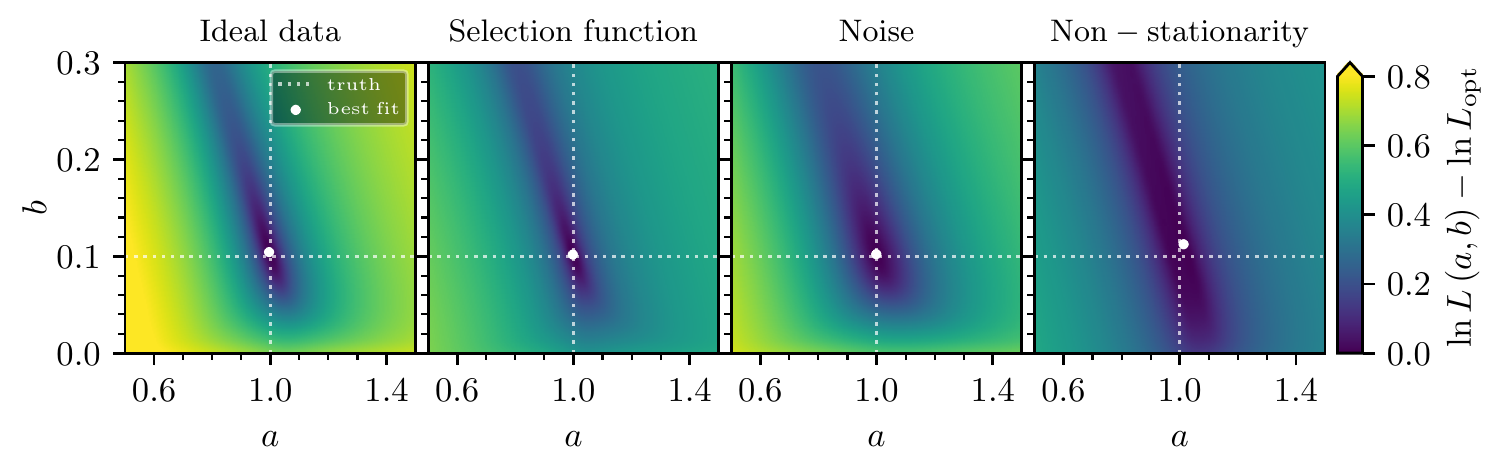}
  \caption{Fitting a simple analytic potential to four different Miyamoto-Nagai mock datasets. From left to right, the panels show the results for mock data with: ideal conditions, an applied selection function, measurement noise, and non-stationarity. For each dataset, we plot the loss $\ln L$ (a measure of the non-stationarity of the distribution function under the assumed potential) as a function of the Miyamoto-Nagai disk potential parameters, $a$ and $b$, computed using the distribution function gradients obtained from our normalizing flow. The true parameter values ($a = 1$, $b = 0.1$) are denoted by white crosshairs, while the parameter values that yield the lowest loss are denoted by a white dot. In all cases, we find close agreement between the true and minimum-loss parameter values, indicating that our normalizing flows provide sufficiently accurate distribution function gradients to tightly constrain the potential, and that the method is robust to various forms of non-ideal data. For ideal mock data, we recover best-fit parameter values of ($a = 0.996$, $b = 0.104$). The largest discrepancy from the true parameter values is found for the non-stationary case, in which our best-fit parameters are ($a = 1.013$, $b = 0.112$).}
  \label{fig:mn-loss-analytic}
\end{figure*}

The Plummer sphere studied above is a highly simplified system, with spherical symmetry and isotropically distributed velocities. In order to study a slightly more complicated system, we consider a population of kinematic tracers orbiting on near-circular orbits in a simple disk potential. Unlike in our previous example, this is not a self-gravitating system. This is, the kinematic tracer particles orbit in a fixed background potential, which is generated by an unseen mass distribution.

We make use of the Miyamoto-Nagai potential \citep{MiyamotoNagai1975}, an axisymmetric disk potential given in cylindrical coordinates by:
\begin{align}
  \Phi \left( R , z \right)
  =
  -\left[
    R^2 + \left( \sqrt{z^2 + b^2} + a \right)^2
  \right]^{-\nicefrac{1}{2}}
  \, .
  \label{eqn:Miyamoto-Nagai-potential}
\end{align}
The Miyamoto-Nagai potential corresponds to a flattened disk, with thickness controlled by the ratio $b/a$. In the limit that $b/a \rightarrow 0$, the potential reduces to the Kuzmin disk \citep{Kuzmin1956,Toomre1963}, which is the potential of a disk with infinitesimal thickness and surface density $\Sigma \left(R\right)$ proportional to $\left( R^2 + a^2 \right)^{-3/2}$. In the limit that $a \rightarrow 0$ and $b/a \rightarrow 0$, the potential reduces to that of the Plummer sphere. We choose $a = 1$, $b = 0.1$, representing a highly flattened disk.

In order to obtain a sample of stationary kinematic tracers, we initialize $2^{19}$ (524,288) particles randomly, and then numerically integrate their trajectories for many dynamical times to allow them to phase-mix. While the initial distribution is non-stationary, the final distribution should be approximately stationary. We draw the initial particle positions from an axisymmetric double-exponential density distribution,
\begin{align}
  \rho \left( R, z \right) \propto \exp \left(
    -\frac{R}{h_R} - \frac{\left|z\right|}{h_z}
  \right)
  \, ,
\end{align}
with $h_R = 1$, $h_z = 0.1$. As the potential is axisymmetric, we do not need to draw the azimuthal angle $\varphi$ in order to integrate orbits. We integrate the particle trajectories in $\left( R, z, v_{\phi}, v_R, v_z \right)$, and then assume that $\varphi$ is uniformly distributed. We initialize the particles on nearly circular orbits, with small initial velocity dispersions in the radial, vertical and transverse (azimuthal) directions. In detail, we draw the three components of the initial velocity as follows:
\begin{align}
  v_R = 0.05 \, v_c \left( R \right) \, \delta_1
  \, , \\
  v_T = v_c \left( R \right) \left( 1 - 0.1 \left| \delta_2 \right| \right)
  \, , \\
  v_z = 0.05 \, v_c \left( R \right) \, \delta_3
  \, ,
\end{align}
where $\delta_i \sim \mathcal{N} \left( 0 , 1 \right)$, and $v_c \left( R \right)$ is the circular velocity in the midplane at the particle's initial orbital radius, $R$. To ensure thorough phase mixing, we integrate each particle for approximately 128 dynamical times. We estimate the dynamical time for each particle as the maximum of three timescales: the orbital, epicyclic and vertical periods. In order to accurately trace the particle trajectories, the integration timestep for each particle is set to $\nicefrac{1}{16^{\mathrm{th}}}$ of the minimum of the preceding three timescales. For the Miyamoto-Nagai disk with our chosen parameters, the longest of the three timescales is always the orbital period, while the shortest timescale is always the vertical period. The particle trajectories are integrated using a 6$^{\mathrm{th}}$-order symplectic method \citep{Yoshida1990Integrator,Kinoshita1991Integrator}, which is implemented by the \texttt{galpy} Python package \citep{Bovy2015galpy}. For more than 99.9\% of the trajectories, energy is conserved to within 0.01\%, indicating that integration errors are negligible.

We feed these particle positions and velocities into our Deep Potential machinery, first learning a normalizing flow representation of the distribution function, and then finding the potential that renders the recovered distribution function stationary. An advantage of our method is that it can be applied to any system for which our assumptions (gravitational dynamics, stationarity and non-negative gravitational mass -- though this latter assumption can be dropped) hold. We thus recover the gravitational potential of our Miyamoto-Nagai disk system using the exact same code, without modification, as we use to recover the potential of the Plummer sphere.

Our resulting gravitational potential and gravitational mass density ($\rho = \nabla^2 \Phi / \left( 4 \pi \right)$) are shown in Fig.~\ref{fig:mn-potential-density}. Fig.~\ref{fig:mn-residuals-xy-plane} compares the potential, gravitational forces ($\vec{F} = -\nabla \Phi$) and gravitational mass density that we recover in the midplane ($z = 0$) of the disk to the true forces felt by particles in a Miyamoto-Nagai potential. We recover the potential to within $\sim$0.1\% and forces to within a few percent. The typical scale of radial force errors in the midplane, $\sim$1\%, corresponds to errors in circular velocity of $\sim$0.5\%. Within the regions of the midplane where $\rho > 0.1 \, \rho_{\mathrm{max}}$, we recover density to within $\sim$10\%. This follows a general trend that we observe across various tests of Deep Potential with mock data: we typically recover the potential to high precision, while precision decreases with each succeeding derivative that we take of the potential (i.e., forces depend on first derivatives of the potential, while density depends on the second derivatives). The right panel of Fig.~\ref{fig:mn-residuals-tracers-Rz} shows the residuals of these same quantities in the $\left(R,z\right)$-plane, averaged over azimuth, while the left panel shows the residuals of gravitational potential, vertical force and density at the locations of the kinematic tracers.

Note that it is also possible to fit simple parameterized models of the potential to the distribution function gradients that are returned by the normalizing flow, by minimizing the loss $L$ (Eq.~\ref{eqn:theta-loss}) w.r.t. the model parameters. We demonstrate this here by fitting a Miyamoto-Nagai potential model (Eq.~\ref{eqn:Miyamoto-Nagai-potential}) to our recovered distribution function gradients. Fig.~\ref{fig:mn-loss-analytic} shows the loss as a function of the Miyamoto-Nagai potential parameters $a$ and $b$. We recover the correct parameters ($a = 1$, $b = 0.1$) with high precision (better than 1\%), demonstrating that our distribution function gradients are sufficiently precise to constrain the potential.

\section{Non-ideal data}
\label{sec:non-ideal}

Most physical systems are non-stationary to some degree, and real-world data is subject to observational errors and selection functions. In this section, we will explore how Deep Potential performs in the presence of these effects. In each case, we use our Miyamoto-Nagai disk model.

\subsection{Observational errors}
\label{sec:observational-errors}

In order to simulate the effects of observational errors, we place a virtual observer at $\left(x, y, z\right) = \left(1,0,0\right)$ in our Miyamoto-Nagai disk, and add independent Gaussian noise to the ``observed'' distances, radial velocities and proper motions. For constant parallax errors, the distance error would scale with the square of distance, $r$, from the observer. We thus add in Gaussian noise with fractional standard deviation given by $\sigma_r / r = \min \left( 0.1\,r, 0.1 \right)$. The fractional distance error grows linearly to 10\% at a distance of $1$, and then remains 10\%. We add in Gaussian proper-motion errors with a standard deviation of 0.0015, regardless of distance. We add in constant proper motion errors, which translate to tangential velocity errors that grow linearly with distance. We choose $\sigma_{v_T} = 0.00015\,r$. These errors are chosen to be similar to what \textit{Gaia} DR3 is expected to achieve for stars with an apparent magnitude of $G = 17$. If our distance units were kiloparsecs, our distance errors would correspond to 0.1~mas (out to $r = 1$). Our radial velocity errors correspond to 1/400$\mathrm{th}$ of the maximum circular velocity (or $\sim 0.5\,\mathrm{km\,s^{-1}}$ in the Milky Way), and our tangential velocity errors are 1/4000$\mathrm{th}$ of the maximum circular velocity (equivalent to measuring proper motion to $\sim 0.1\,\mathrm{mas\,yr^{-1}}$ at a distance of 1~kpc in the Milky Way).

The accuracy of our recovered potential, vertical forces and densities are summarized in the middle panel of Fig.~\ref{fig:mn-non-ideal-residuals-tracers}. For the level of noise assumed in this test, the force and density residuals are only slightly larger than what we obtain with ideal mock data. This indicates that we can expect good results for the Milky Way using \textit{Gaia} DR3 data for stars with apparent magnitudes  down to an apparent magnitude of $G \sim 17$. In order to push into the low-signal-to-noise regime, however, we expect that deconvolution would be required to more accurately recover the true distribution function.

\subsection{Selection functions}
\label{sec:selection-functions}

\begin{figure*}
  \centering
  \includegraphics[width=\linewidth]{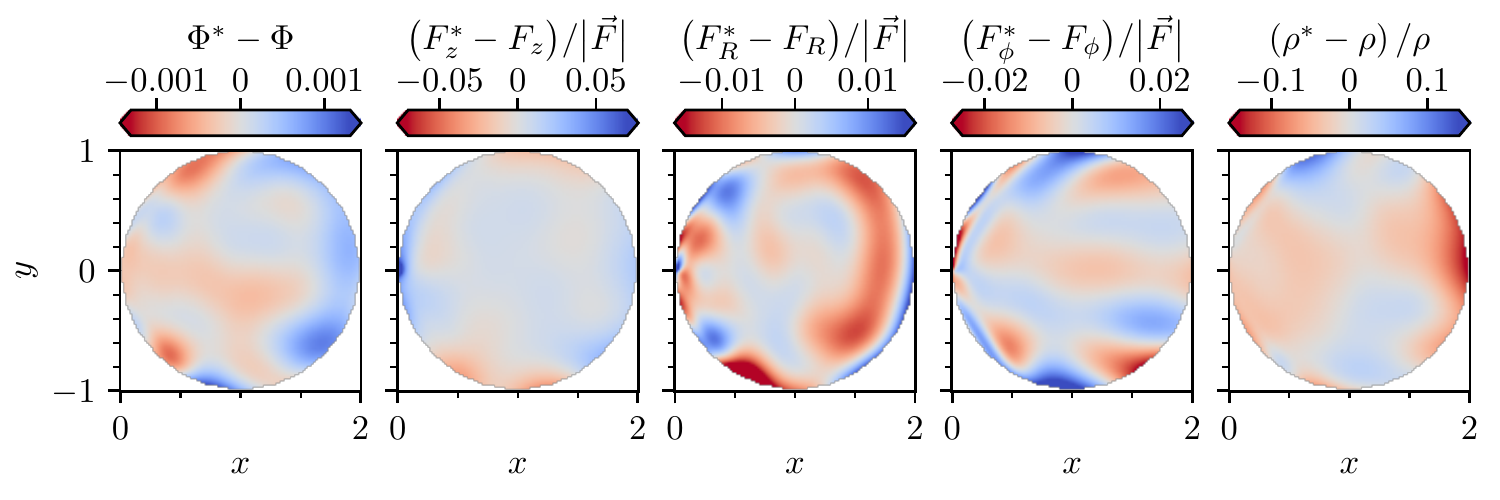}
  \caption{As Fig.~\ref{fig:mn-residuals-xy-plane}, but for our mock dataset with an applied selection function. The unobserved regions of the disk midplane are shown in white. In the region of the Miyamoto-Nagai-disk midplane with mock data, we obtain the potential, forces and densities to a similar accuracy as in our mock dataset with no selection function.}
  \label{fig:mn-sel-residuals-xy-plane}
\end{figure*}

\begin{figure*}
  \centering
  \begin{minipage}{0.32\linewidth}
      \centering
      \hspace{1.0cm}{With selection function}\par\medskip
      \vspace{-0.25cm}
      \includegraphics[width=\linewidth]{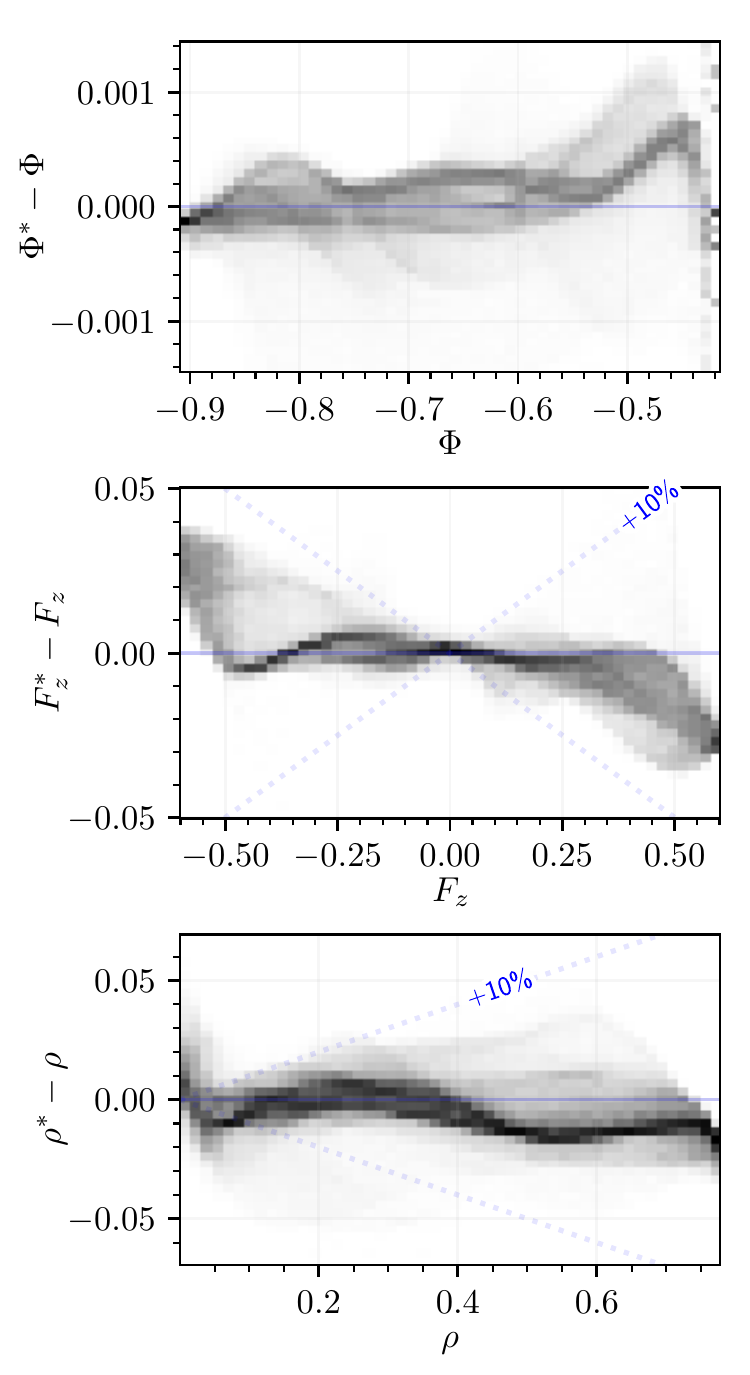}
  \end{minipage}
  \begin{minipage}{0.32\linewidth}
      \centering
      \hspace{1.0cm}{With noisy data}\par\medskip
      \vspace{-0.25cm}
      \includegraphics[width=\linewidth]{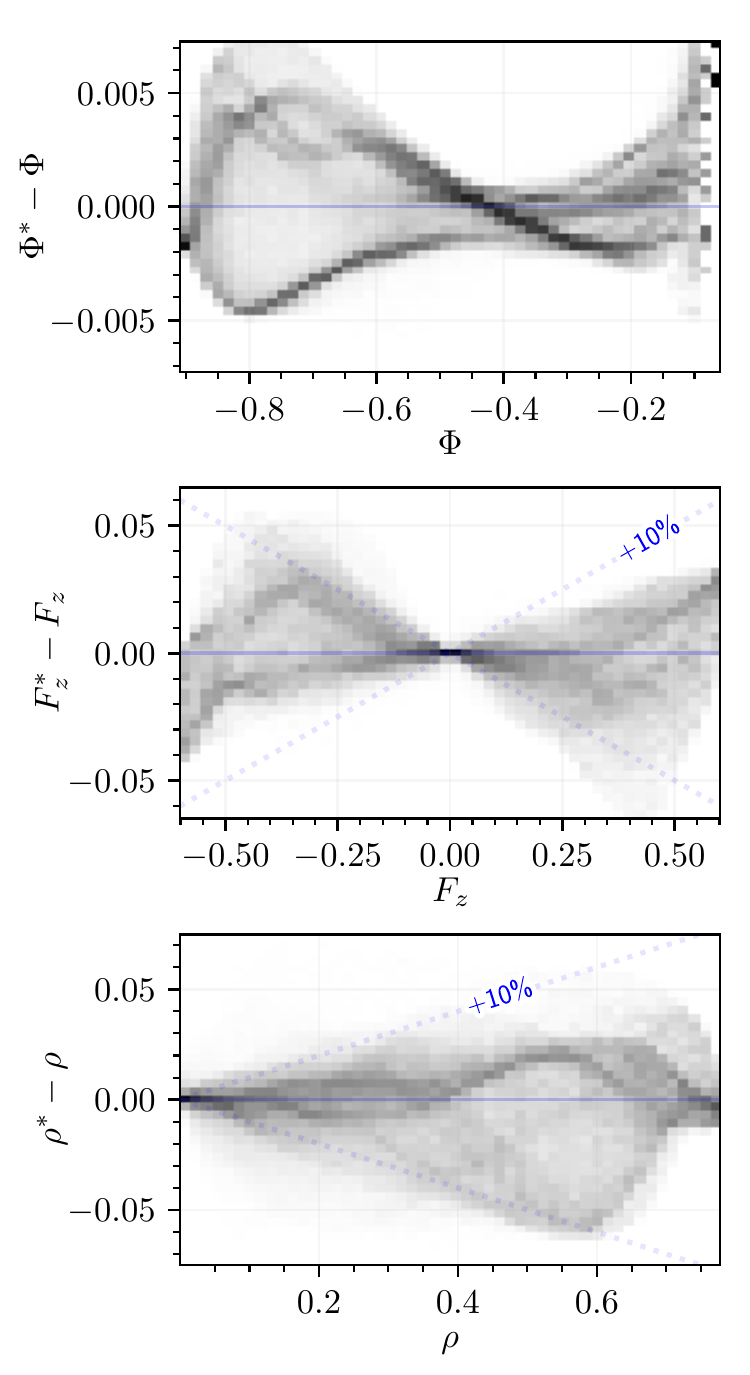}
  \end{minipage}
  \begin{minipage}{0.32\linewidth}
      \centering
      \hspace{1.0cm}{With non-stationarity}\par\medskip
      \vspace{-0.25cm}
      \includegraphics[width=\linewidth]{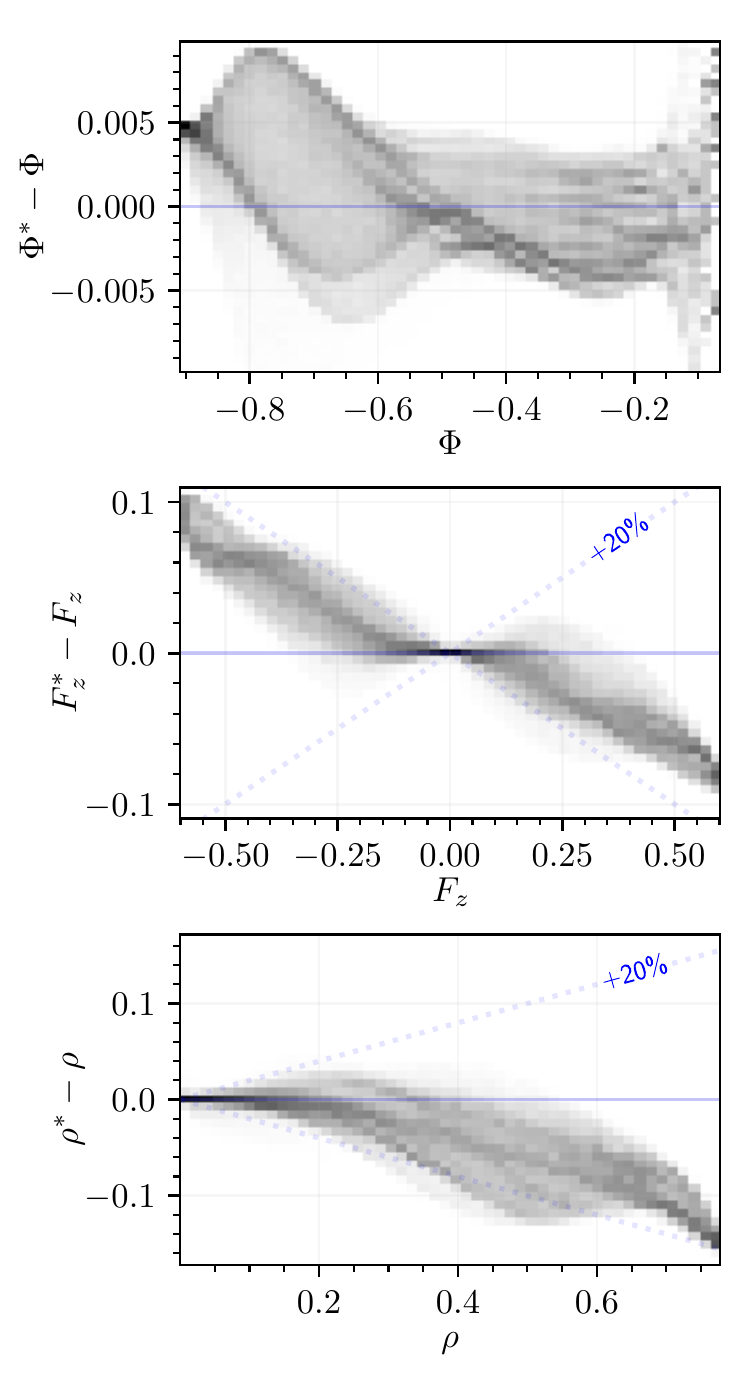}
  \end{minipage}
  \caption{As the left panel of Fig.~\ref{fig:mn-residuals-tracers-Rz}, but for our mock dataset with an applied selection function (left column), with noisy data (middle column) and with non-stationarity (right column). As can be seen in the left column, in the region of the Miyamoto-Nagai-disk with mock data, we obtain the potential to an accuracy of $\sim$0.1\%, and the vertical restoring force and densities to better than 10\%. Introducing measurement noise into the data increases the residuals slightly, but vertical forces and densities are still measured to within $\sim$10\%. The introduction of significant non-stationarity into the system has a larger effect, introducing a bias into the recovered vertical forces and densities of slightly less than 20\%. Nevertheless, the method recovers the potential, forces and densities remarkably well, given the sharp caustics in the non-stationary mock datasets.}
  \label{fig:mn-non-ideal-residuals-tracers}
\end{figure*}

As currently formulated, our method does not take into account selection functions. Nevertheless, because Deep Potential only uses information about local distribution function gradients to determine the gradient of the gravitational potential at each point, Deep Potential can be applied to observational data for which the selection function is a constant within some volume of phase space. In the volume in which the selection function is constant, the recovered distribution function will be unaffected (up to a normalizing constant), and the gravitational potential that renders the system stationary will be unchanged.

It is common for observational selection functions to depend on position (and often luminosity and color), but not on velocity. We therefore apply a simple selection function to our Miyamoto-Nagai disk mock data: all stars within a sphere of unit radius centered on $\left( x, y, z \right) = \left( 1, 0, 0 \right)$ are observed, while all stars outside of this sphere are unobserved. We generate a larger amount of mock data, so that after application of the selection function, $2^{19}$ particles (the same number as in our other mock datasets) remain. We apply Deep Potential to this mock dataset. As demonstrated in Figs.~\ref{fig:mn-sel-residuals-xy-plane} and \ref{fig:mn-non-ideal-residuals-tracers}, within the volume in which we have data, we are able to recover the potential, forces and density to similar accuracy as in the ideal case.

When the selection function is not constant within the observed volume, the method laid out in this paper would require modification. There are two broad approaches to dealing with non-uniform selection functions. The first approach is to take the selection function into account when training normalizing flow that represents the distribution function, in order to recover the true underlying distribution function (corrected for selection effects). The second approach is to modify the stationarity condition, Eq.~\eqref{eqn:stationarity}, to include the selection function. We leave these considerations to future work, and suggest that initial real-world applications of Deep Potential should rely on carefully constructed datasets of kinematic tracers with selection functions that are approximately constant within some volume.

\subsection{Non-stationarity}
\label{sec:non-stationarity}

\begin{figure*}
  \centering
  \setlength{\lineskip}{-7.25pt}
  \includegraphics[width=0.88\linewidth]{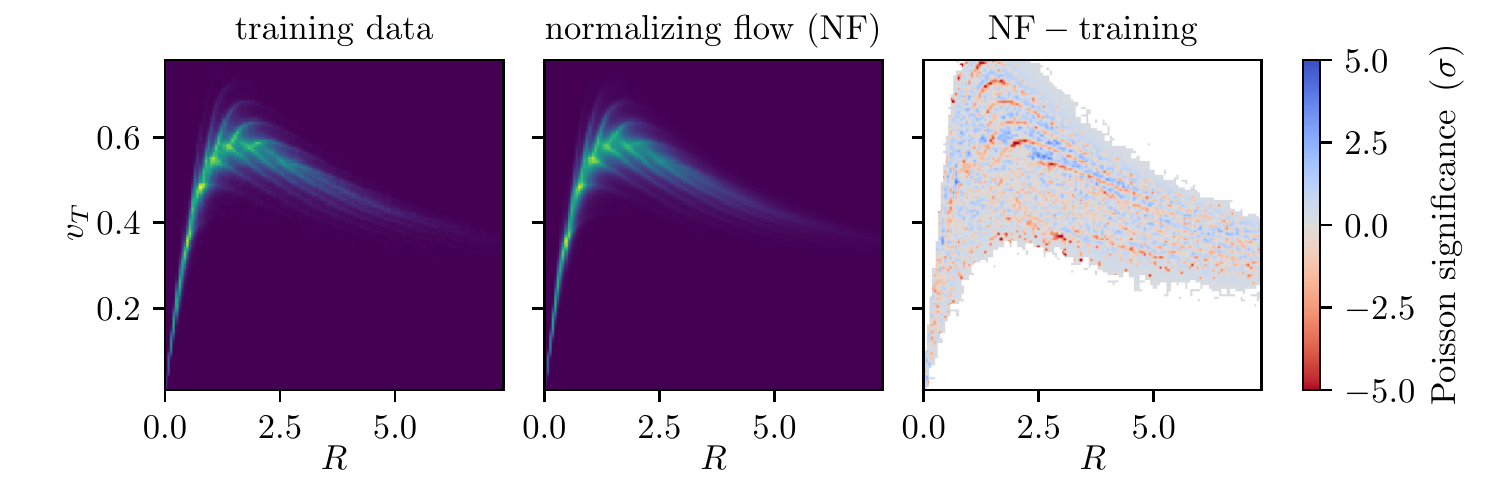}
  \includegraphics[width=0.88\linewidth]{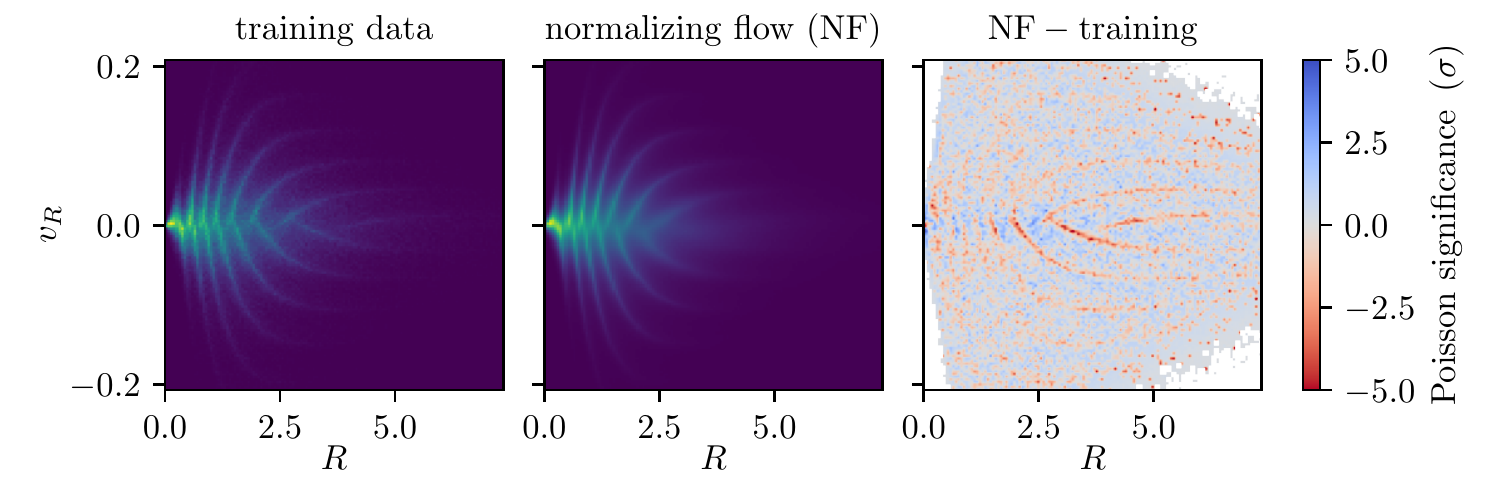}
  \includegraphics[width=0.88\linewidth]{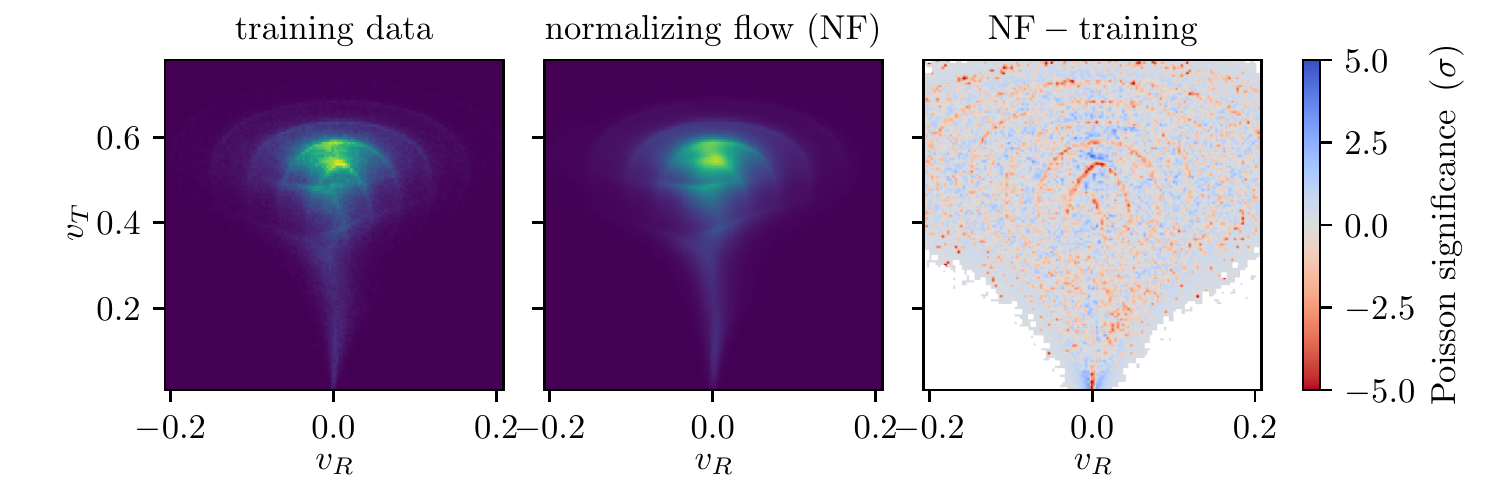}
  \includegraphics[width=0.88\linewidth]{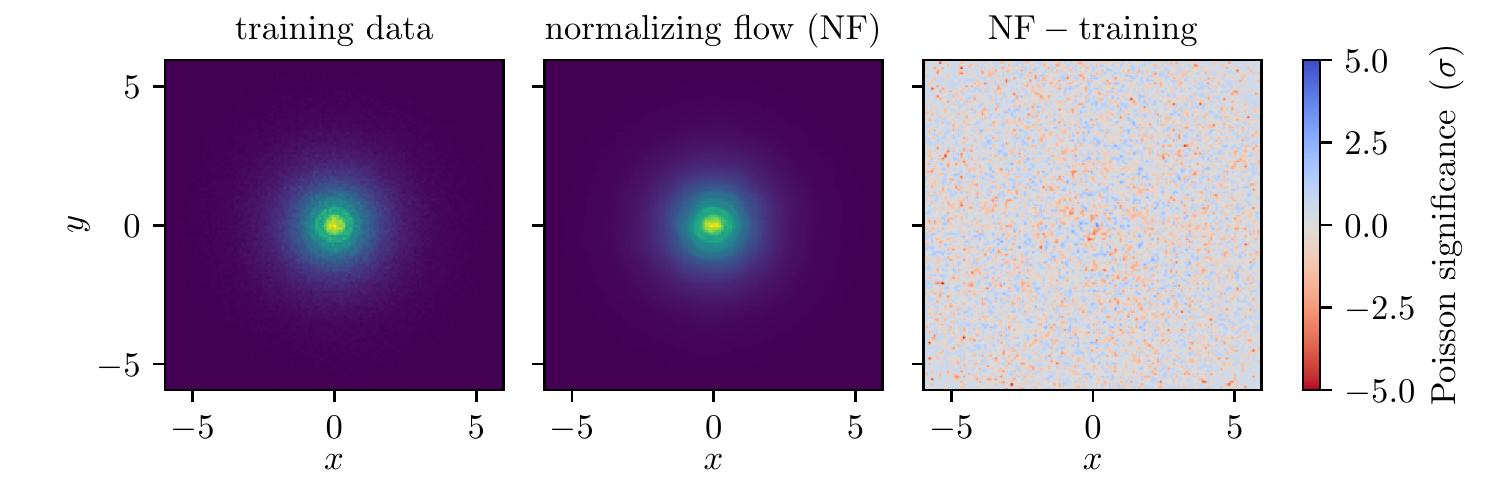}
  \caption{Comparison of the distribution of the input training data for the non-stationary Miyamoto-Nagai disk (left column) with our normalizing flow (middle column) in several different two-dimensional projections of phase space. The right column shows the Poisson significance of the differences between the training data and our normalizing flow. Even for this highly challenging dataset, which contains many sharp caustics in phase space, our normalizing flow performs fairly well, recovering most of the sharp edges in the distribution function. This test is designed to represent an extreme case of non-stationarity, and the actual Milky Way stellar distribution function is expected to be much smoother than this test case.}
  \label{fig:mn-nonstationary-df}
\end{figure*}

\begin{figure}
  \centering
  \includegraphics[width=0.45\textwidth]{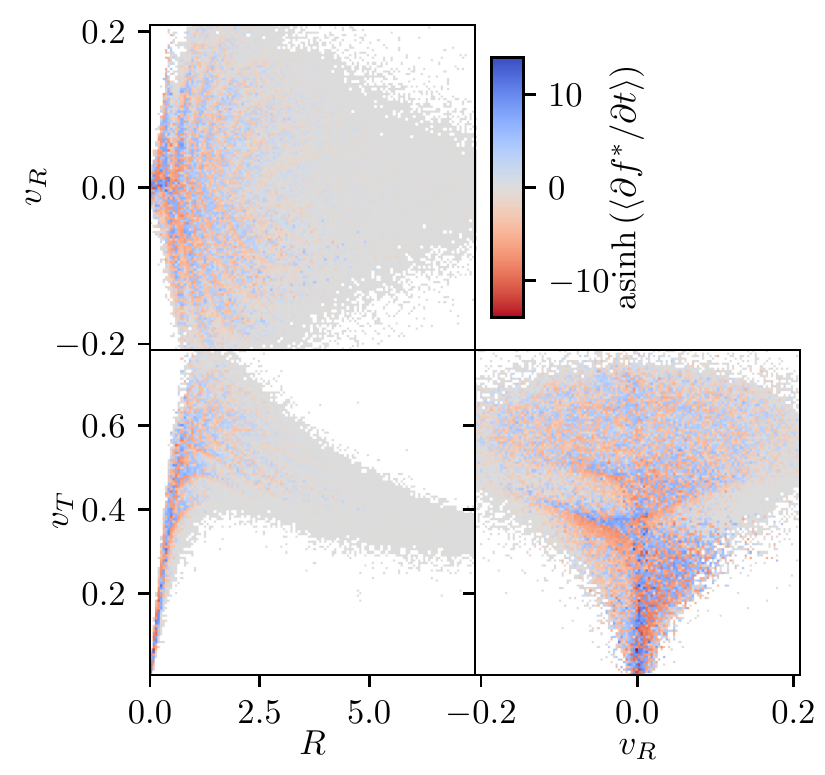}
  \caption{Visualizing non-stationarity in phase space in our non-stationary Miyamoto-Nagai disk. In each panel, we plot a projection of the rate of change of the distribution function (averaged over the remaining phase-space dimensions, weighted by the distribution function) implied by our solution for the distribution function and potential. We compress the color scale using an $\mathrm{asinh}$ function, so that both small and large variations are visible.}
  \label{fig:mn-nonstationarity}
\end{figure}

\begin{figure*}
  \centering
  \includegraphics[width=0.95\linewidth]{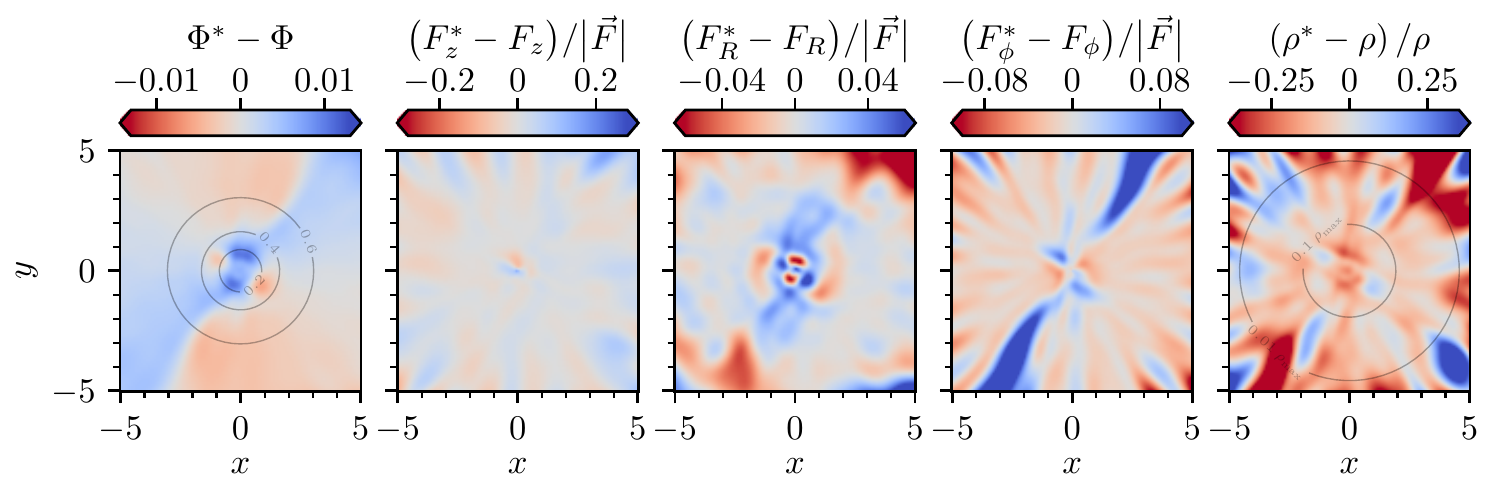}
  \caption{As Fig.~\ref{fig:mn-residuals-xy-plane}, but for our non-stationary Miyamoto-Nagai disk mock dataset. Despite the presence of sharp caustics in phase space, in the midplane of the disk, we are able to recover the potential to $\sim$1\%, forces to better than 10\%, and densities to $\sim$20\%.}
  \label{fig:mn-nonstat-residuals-xy-plane}
\end{figure*}

The Milky Way is not a perfectly stationary system. The Galactic bar and spiral arms are inherently non-stationary, the disk is perturbed by infalling satellite galaxies, such as the Sagittarius dwarf galaxy \citep{Laporte2019Sgr}, and dynamical times in the Galactic halo are on the order of Gigayears. In the Galactic disk, in particular, \textit{Gaia} has revealed non-stationary features, such as a phase spiral in the $\left( z, v_z \right)$-plane \citep{Antoja2018Nature}. It is thus important to test the performance of Deep Potential in systems with some amount of non-stationarity. In such systems, rather than finding a potential which renders the distribution system stationary (and in most non-stationary systems, no such potential exists), we find the potential that minimizes our measure of non-stationarity.

In order to obtain a slightly non-stationary system, we truncate our integration of particle trajectories in the Miyamoto-Nagai disk after only 8 dynamical times (rather than the usual 128 dynamical times), before they have fully phase-mixed. Fig.~\ref{fig:mn-nonstationary-df} compares the distribution function of the particles at 8 and 128 dynamical times.

Despite the presence of sharp caustics in phase space, we are nevertheless able to recover the gravitational potential to $\sim$1\%, forces to within 10\%, and densities to within $\sim$20\%, as seen in Fig.~\ref{fig:mn-nonstat-residuals-xy-plane}. The surprising robustness of Deep Potential in the presence of severe non-stationarity is likely due to the fact that at each location in space, a single force vector has to minimize non-stationarity over all velocities. At one point in physical space, different regions of velocity space may ``prefer'' different force vectors, with the errors introduced by different regions of velocity space partially canceling one another out. The possibility that such cancellation occurs can be seen more clearly in Fig.~\ref{fig:mn-nonstationarity}, which visualizes non-stationarity in three different projections of phase space. For example, holding cylindrical radius fixed but varying $v_T$, $\partial f / \partial t$ rapidly oscillates between positive and negative values, indicating inflow or outflow of particles, respectively. Neighboring inflows and outflows tend to pull the optimal solution for the local force vector in opposite directions, partially cancelling one another's effect on the solution to the potential.

\section{Discussion}
\label{sec:discussion}

Deep Potential is a fairly direct approach to solving the Collisionless Boltzmann Equation: we first model the observed distribution function, and then directly search for a potential that minimizes non-stationarity. The question then arises: why hasn't this approach been attempted previously? We believe there are two primary reasons. First, our approach makes use of relatively new tools from machine learning. In particular, normalizing flows give us the ability to learn the underlying smooth (and possibly highly complicated) probability distribution function of a set of observed points. Second, and perhaps more importantly, this approach relies on the availability of large numbers of stars with measured six-dimensional phase-space coordinates. Prior to \textit{Gaia} and large ground-based spectroscopic surveys, such as LAMOST and APOGEE \citep{Majewski2017APOGEE}, catalogs of stars with six-dimensional phase-space information were insufficient to model the full distribution function in a model-free manner, and parameterized models with relatively small numbers of adjustable parameters were required. The ongoing dramatic increase in the available amount of phase-space data allows us to move beyond simple parametric models, and to discard some of the assumptions that often underlie them (such as axisymmetry).

Going beyond simplifying assumptions about the distribution function is particularly important at the present moment, as \textit{Gaia} has shown us that the true distribution function of the Milky Way is rich in details. Deep Potential makes use of the whole distribution function, rather than just broad features, and therefore more fully extracts the information that is available in phase space. The output of Deep Potential is not simply the gravitational potential: after learning the potential, we can look for evidence of non-stationary features in the distribution function. On real datasets, we expect diagnostics such as Fig.~\ref{fig:mn-nonstationarity}, which shows the rate of inflow or outflow in each region of phase space, to be useful in identifying and mapping non-stationary features.

Deep Potential makes few model assumptions, the most significant being that the dynamical system is stationary. However, as we demonstrate with our non-stationary Miyamoto-Nagai mock dataset in Section~\ref{sec:non-stationarity}, Deep Potential is robust to non-stationary caustics in the distribution function that are larger than what we expect to observe in the Milky Way. Because it makes few model assumptions, Deep Potential can be applied to any system with six-dimensional phase-space measurements with minimal modification. There is no need to choose analytic forms for distribution function and potential, or to specify the types of spatial symmetris expected in the problem. In this paper, we have demonstrated that the same pipeline works for a spherically symmetric system (Section~\ref{sec:plummer-sphere}), an axisymmetric Miyamoto-Nagai disk (Section~\ref{sec:miyamoto-nagai}), a system with a spatial selection function with sharp boundaries (Section~\ref{sec:selection-functions}), and a system with significant non-stationarity (Section~\ref{sec:non-ideal}).

There are several future directions in which Deep Potential can be developed. First, our current formulation can only handle selection functions that are constant in the region of interest. In Section~\ref{sec:selection-functions}, we impose a selection function which is constant within a spherical volume, and zero outside. However, a selection function that varies smoothly within the region of interest would require a modification of our formalism.

Second, our formalism does not include observational errors. In Section~\ref{sec:observational-errors}, we show that in a toy model, small observational errors do not significantly degrade our recovered gravitational potential. However, accommodation of observational errors in the formalism would allow us to apply this method to data with less precisely measured phase-space information (in particular, less precise parallaxes and proper motions).

Third, our current implementation of Deep Potential fixes the frame of reference in which the system is stationary. As laid out in Section~\ref{sec:additional-considerations}, it is possible to introduce additional free parameters that encode a change of reference frame. This could, for example, accommodate systems that are stationary in a rotating frame, or in a frame that is moving relative to the observer. This change of frame may itself have physical meaning, as it may indicate the rest frame of the Galactic barycenter, or the pattern speed of the bar or spiral arms.

Fourth, we assume that all six phase-space dimensions are observed. With \textit{Gaia}, six-dimensional phase-space data will be available in much of the Milky Way. However, in external galaxies, or in very compact systems (such as globular clusters), we will generally lack at least one dimension (e.g., line-of-sight distance within the system). One approach to dealing with missing dimensions is to impose symmetries on the system, though we leave this line of investigation to future work.

Fifth, stellar metallicity offers an entirely new dimension that could be incorporated into Deep Potential. As pointed out in \citet{PriceWhelan2021OribtalTorusImaging}, for phase-mixed systems, the stellar metallicity distribution should be constant along orbits. \citet{PriceWhelan2021OribtalTorusImaging} develops a method for finding the gravitational potential that aligns stellar orbits with surfaces in phase space of constant metallicity. This suggests a possible extension of Deep Potential: imposing stationarity on the distribution function, \textit{conditional on metallicity}. For example, one could model the conditional distribution function, $f \left( \vec{x}, \vec{v} \mid \left[\mathrm{Fe}/\mathrm{H}\right] \right)$, as a normalizing flow, and then find the potential that renders it stationarity at a large number of points drawn from the joint distribution $p \left( \vec{x}, \vec{v}, \left[\mathrm{Fe}/\mathrm{H}\right] \right)$. We likewise leave development of such an approach to future work.

Finally, we would like to point to a related, likewise promising method, developed by \citet{An2021Uniqueness} and \citet{Naik2022}, which similarly builds on \citet{GreenTing2020NeurIPS}. This closely related method similarly uses a normalizing flow to recover the gradients of the distribution function. It differs from our method primarily in that it directly solves for accelerations -- rather than the potential field -- at discrete points in space, based on a least-squares minimization of a measure of non-stationarity. The most significant difference between these two approaches, in our view, is that by solving for the potential field $\Phi \left(\vec{x}\right)$, one can place a positivity constraint on the underlying matter density ($\nabla^2 \Phi \geq 0$) and can penalize overly complicated potentials (e.g., through $L_2$ weight regularization). In the final stages of the preparation of the present manuscript, \citet{Buckley2022GalacticDarkMatterML} published a related technique, which derives accelerations using a similar method to \citet{An2021Uniqueness}, and then calculates densities using Gauss' theorem.

\section{Conclusions}
\label{sec:conclusions}

In this paper, we have shown that it is possible to accurately recover the gravitational potential of a stationary system using a snapshot of a sample of phase-mixed tracers. Rather than using moments or simple parametric models of the distribution function, our method makes use of the whole distribution function, in all its complexity. Auto-differentiation computational frameworks such as Tensorflow and PyTorch provide ideal tools for creating smooth and differentiable -- yet highly flexible -- representations of the distribution function and gravitational potential. Our method, ``Deep Potential,'' first represents the distribution function using a normalizing flow, and then finds the gravitational potential -- represented using a feed-forward neural network -- that renders the distribution function stationary.

Making use of the full distribution function is all the more important now, as orders of magnitude more six-dimensional phase-space measurements of Milky Way stars are just now becoming available. The \textit{Gaia} space telescope is currently surveying parallaxes and proper motions of over a billion stars, and is additionally measuring radial velocities of tens of millions of stars \citep{Prusti2016}. Ground-based spectroscopic surveys are set to deliver millions more high-precision radial velocities over the coming years \citep{Steinmetz2006,Cui2012LAMOST,Zhao2012LAMOST,DeSilva2015GALAH,Majewski2017APOGEE,Kollmeier2017}. We thus will soon have access to precise six-dimensional phase-space coordinates of tens of millions of stars throughout a large volume of the Milky Way.

Deep Potential provides a means of extracting the gravitational potential -- and therefore the three-dimensional distribution of baryonic and dark matter in the Galaxy -- from the rich kinematic datasets that are now becoming available. Deep Potential makes only minimal physical assumptions (steady-state dynamics in a background gravitational field that corresponds to a positive matter density), without resorting to restricted analytical models of the distribution function and potential. We have demonstrated that Deep Potential performs well on ideal mock data, and have demonstrated its robustness in the presence of non-stationarity, observational errors and simple selection functions. This method therefore has the potential to reveal the full, unseen mass distribution of the Galaxy, using only a set of visible kinematic tracers -- the stars.

\section{Acknowledgments}

Gregory Green acknowledges funding from the Alexander von Humboldt Foundation, through the Sofja Kovalevskaja Award. Yuan-Sen Ting acknowledges financial support from the Australian Research Council through DECRA Fellowship DE220101520.

The authors thank Hans-Walter Rix for helpful discussions during hikes up the K\"{o}nigstuhl in Heidelberg; Stephen Portillo, Douglas Finkbeiner and Joshua Speagle for initial discussions at ``Cape Code,'' where the idea for the method first came into being; and Wyn Evans for discussions of the method and alternative approaches over email and Zoom.


\bibliography{bibliography}{}
\bibliographystyle{aasjournal}



\appendix

\section{Existence and uniqueness of stationary solutions to the Collisionless Boltzmann Equation}
\label{app:existence-uniqueness}

In Section~\ref{sec:method}, we noted that in almost all physically plausible stationary systems, there is a unique solution for the gravitational potential (up to an additive constant). Here, we show that \textit{local gradients} of the gravitational potential are fully determined by the stationarity condition.

At a fixed location in space, $\vec{x}_0$, consider the behavior of the distribution function as the velocity $\vec{v}$ varies. The stationarity condition, Eq.~\eqref{eqn:stationarity}, takes the form,
\begin{align}
    \vec{a} \cdot \nabla_{v} f \left( \vec{x}_0, \vec{v} \right)
    &=
    -\vec{v} \cdot \nabla_x f \left( \vec{x}_0, \vec{v} \right) \, ,
    \label{eqn:stationarity-at-one-x}
\end{align}
where $\vec{a} = -\nabla \Phi$ is the gravitational acceleration a test mass would feel at $\vec{x}_0$. In general, in a three-dimensional system, if we measure the gradients of the distribution function $f$ at three different velocities, then we have three unknowns (the three components of $\vec{a}$) and three equations, and the local acceleration should be fully specified. If we measure the gradients of the distribution function at many different velocities, then in general, we should expect the system to be overdetermined. Stationarity is thus a very strict condition, which implies a certain relation between the spatial and velocity gradients of the distribution function as we vary velocity, holding position fixed.

We can put this line of thought on a somewhat firmer footing. Imagine measuring the gradients of the distribution function at $N$ velocities, $\vec{v}_1, \ldots, \vec{v}_N$, while holding position, $\vec{x}_0$, fixed. Define a matrix whose rows are the velocity gradients of the distribution function at the different observed velocities,
\begin{align}
    \mathbf{B} &\equiv
    \left(
        \nabla_v f \left(\vec{x}_0,\vec{v}_1\right) ,\,
        \cdots ,\,
        \nabla_v f \left(\vec{x}_0,\vec{v}_N\right)
    \right)^T
    \, ,
\end{align}
and a vector related to the spatial gradients of the distribution function,
\begin{align}
    c &\equiv
    -\begin{pmatrix}
        \vec{v_1} \cdot \nabla_x f \left(\vec{x}_0,\vec{v}_1\right) \\
        \vdots \\
        \vec{v_N} \cdot \nabla_x f \left(\vec{x}_0,\vec{v}_N\right)
    \end{pmatrix} \, .
\end{align}
Then, the stationarity condition (Eq.~\ref{eqn:stationarity-at-one-x}) at the set of observed velocities can be written as a linear system of equations:
\begin{align}
    \mathbf{B} \, \vec{a} &= \vec{c} \, .
\end{align}
The Rouch\'{e}-Capelli theorem then tells us that if the rank of $\mathbf{B}$ is equal to the dimensionality of $\vec{a}$ (i.e., the number of spatial dimensions in the system), then there is at most one solution to the linear system of equations. This is the case if the velocity gradients, $\nabla_v f \left(\vec{x},\vec{v}_i\right)$, $i \in 1, \ldots, N$, span the entire space (i.e., $\mathbb{R}^3$ for 3-dimensional space). Holding $\vec{x}$ fixed and varying $\vec{v}$, the distribution function $f \left(\vec{x}, \vec{v}\right)$ must be normalizable (that is, the probability density function of velocity at a given position must integrate to unity). If it is also continuous in $\vec{v}$ (again, holding $\vec{x}$ fixed), then the set of its velocity gradient vectors at all values of $\vec{v}$ spans the entire space. Given observations of the distribution functions at a sufficient number of different velocities, the local acceleration vector $\vec{a}$ (and thus the local gradient of the gravitational potential) will be uniquely determined, or will have no solution. Whether or not a solution exists then depends on whether the rank of the augmented matrix $\left[ \mathbf{B} \mid \vec{c} \, \right]$ is equal to the dimensionality of the system. However, if the distribution function is discontinuous, then it is possible for its velocity gradients to not span the full space, and for there to be an infinite number of solutions for $\vec{a}$. In plausible physical systems, however, the distribution function will be continuous, and there will thus be at most one solution for $\vec{a}$ at any point in space.

In general, we expect there to be \textit{no solution} for the local acceleration $\vec{a}$. Even if the underlying physical system is perfectly stationary, we will generally be dealing with noisily measured distribution function gradients. Unless the measurement noise has very particular properties, the rank of the augmented matrix, $\left[ \mathbf{B} \mid \vec{c} \, \right]$, will be greater than the dimensionality of the system. The best we realistically can hope to do, therefore, is to \textit{minimize} some measure of the non-stationarity of the system. This is the approach we take in this work.

\end{CJK*}
\end{document}